\documentclass[%
 reprint,
superscriptaddress,
longbibliography,
 amsmath,amssymb,
 aps,
prb,
]{revtex4-1}
\setcitestyle{numbers,square}

\usepackage{braket}
\usepackage{graphicx}% Include figure files
\usepackage{dcolumn}% Align table columns on decimal point
\usepackage{bm}% bold math
\usepackage[normalem]{ulem} % strikethrough
\usepackage{color}

\usepackage{hyperref}% add hypertext capabilities
\hypersetup{
    colorlinks,%
    citecolor=blue,%
    linkcolor=blue,%
    urlcolor=blue
}

\newcommand{\orcid}[1]{\href{https://orcid.org/#1}{\includegraphics[width=8pt]{orcid.pdf}}}

\begin{document}

\title{Stacking order effects on the energetic stability and electronic properties of $n$-doped graphene/h-BN van der Waals heterostructures on SiC(0001)}

\author{D. P. de Andrade Deus}
\email{dominike@iftm.edu.br}
\affiliation{Instituto Federal de Educa\c{c}\~ao, Ci\^encia e Tecnologia do Tri\^angulo Mineiro, Uberaba, C. P. 1020, 38064-190, Minas Gerais, Brazil}
\author{J. M. J. Lopes}
%\email{lopes@pdi-berlin.de}
\affiliation{Paul-Drude-Institut für Festk\"orperelektronik, Leibniz-Institut im Forschungsverbund Berlin e.V., 10117 Berlin, Germany}
\author{Roberto H. Miwa}
\email{hiroki@ufu.br}
\affiliation{Instituto de F\'isica, Universidade Federal de Uberl\^andia,        C.P. 593, 38400-902, Uberl\^andia, MG, Brazil}

\date{\today}
  
\begin{abstract}
\vspace{3mm}
\begin{center}
 {\bf ABSTRACT}
\end{center}
Heterostructures based on the stacking of two-dimensional materials with different electronic properties have been the subject of several studies addressing the development of novel devices with multiple functionalities. Among them, van der Waals (vdW) systems composed of graphene (G) and hexagonal boron nitride (h-BN) have been extensively investigated, including recent experimental studies on the large-scale synthesis of h-BN on graphene/SiC(0001) templates. Interestingly, their results suggest that besides vdW epitaxy of h-BN on graphene, confined growth of h-BN monolayers between graphene and SiC(0001) could also take place, leading to h-BN encapsulation. In this work, we have performed a theoretical study, based on first-principles calculations, of G/h-BN heterostructures on SiC(0001), the latter covered by a carbon buffer layer (SiC).  Our findings reveal that, indeed, there is an energetic preference for the encapsulation of h-BN below a single layer of graphene on SiC, viz.: G/h-BN/SiC in the case of bilayer systems and G/h-BN/G/SiC in the case of trilayer systems. Electronic structure calculations show that the linear energy band dispersion of graphene is preserved in the bilayer systems independent of the stacking order, however, with the Dirac points lying in different energy positions due to the different electron doping level of graphene. In the trilayer systems, the $n$-type doping level of graphene also depends on the stacking order. The electronic band structure of G/h-BN/G/SiC is characterized by two Dirac points, both below the Fermi level, but with different energies, while the other (energetically less stable) systems, G/G/h-BN/ and h-BN/G/G/SiC are characterized by the emergence of parabolic bands near the Fermi level.  Additional structural characterizations of these G/h-BN heterostructures on SiC were carried out based on simulations of C-1s core-level-shift (CLS) and carbon K-edge X-ray absorption near edge spectroscopy (XANES), with the goal of assisting future experimental spectroscopy in these graphene/h-BN vdW systems.
\end{abstract}

\maketitle

\section{Introduction}

Van der Waals (vdW) heterostructures combining different two-dimensional (2D) materials are very promising for the realization of atomically thin devices with tailored properties and functionalities\cite{geim2013van}. No matter what combination of 2D crystals is desired, their fabrication is usually achieved via sequential stacking of individual flakes exfoliated from bulk crystals \cite{geim2013van,massicotte2016photo}. Although this method allows for preparing samples and devices with a high degree of perfection and quality, the small lateral size (tens of $\mu\text{m}$) of the produced material is unsuitable for applications. Moreover, interface contamination introduced during heterostructure assembly \cite{haigh2012cross,rooney2017observing} may deteriorate the resulting properties. 

A promising alternative to such method is to realize the large-scale synthesis of 2D materials on top of each other via vdW epitaxy \cite{koma1985fabrication}, utilizing either physical- or chemical-vapour deposition methods. Nevertheless, the controlled realization of single crystalline heterostructures via vdW epitaxy is challenging due to additional effects that may take place besides vdW growth. For example, the existence of surface defects such as grain boundaries and nanoholes results in the intercalation of adatom species underneath the 2D template, which may lead to the formation of another material (2D or 3D) at the interface between the 2D template and the host substrate. Feldberg et al.\cite{feldberg2019spontaneous} reported on the spontaneous intercalation of Ga species during growth of GaN on epitaxial graphene on SiC(0001) via molecular beam epitaxy (MBE). Specifically, they observed the formation of an ordered Ga bilayer between the carbon buffer layer and SiC(0001). Similarly, the results from Heilmann et al. \cite{heilmann2DMat2018defect} for MBE growth of h-BN on epitaxial graphene/SiC(0001) suggests that adatom intercalation below graphene takes place.  This result is in agreement with their most recent study \cite{heilmann2021spatially}, in which epitaxial graphene [on SiC(0001)] having a defect patterning produced by He ions was utilized as a growth template for h-BN. In this case, besides the nucleation (and lateral growth) of insulating h-BN monolayer islands on the surface (as unambiguously confirmed by conductive atomic force microscopy), the single-layer graphene regions also exhibited a roughened and somewhat wrinkled surface morphology that was however still conductive, as expected for graphene. This suggests that the growth of another material, possibly boron nitride, took place below single-layer graphene, mostly likely due to the B and N intercalation. It is worth noting that intercalation has also been observed during epitaxial growth of h-BN/graphene heterostructures on metals.\cite{wang2021epitaxial,yang2015creating}

If, on the one hand, intercalation during vdW epitaxy is unwanted, it also represents a very promising path for the controlled synthesis of novel material systems. In fact, Al Balushi et al. \cite{al2016two} developed an experimental scheme that enhanced the intercalation of Ga and N species below epitaxial graphene on SiC(0001), which allowed for the realization of a 2D-like GaN phase encapsulated between graphene and SiC(0001). This approach has recently been generalized for achieving confined epitaxy of 2D metals as well as oxides at the interface between graphene and SiC(0001) \cite{lee2022confined,turker20232d}. It could in principle also be implemented to controllably create an atomically thin h-BN below epitaxial graphene, which could lead to the electronic decoupling of graphene from the buffer layer/SiC(0001).  

Considering the reported facts, we performed a theoretical study, based on first-principles calculations, of the energetic stability and the structural and electronic properties of h-BN/graphene heterostructures on SiC(0001).  Our findings confirm that 2D heterostructures where h-BN is encapsulated between graphene and buffer layer/SiC, or two graphene layers (i.e. graphene/h-BN/graphene on buffer layer/SiC), are energetically favorable when we consider bilayer (BL) and trilayer (TL) systems, respectively. Further calculations revealed that the graphene/h-BN stacking order on SiC(0001) can be characterized by different fingerprints on the electronic band structure and net charge transfer (i.e. electron and hole doping) profiles through the heterostructure. Finally, we have performed a set of spectroscopic simulations, carbon K-edge X-ray absorption near edge structure (XANES), and C-$1s$ core-level-shift (CLS) in order to provide structural information about these stacked layers combined with the local electronic properties.

%\textcolor{red}{Very recently, Heilmann {\it et al.} performed the epitaxial growth of 2D graphene/boron-nitride vdW heterostructures mediated by the creation of spatially ordered defects (vacancies) on the graphene/SiC(0001) substrate.\cite{heilmann2DMat2018defect,heilmannACSApplMat2020influence,heilmann2021spatially} Those defects act as nucleation sites for the  growth of 2D h-BN sheets \textcolor{red}{at (embedded) the (in) G/SiC interface (?).} Indeed, defect engineering has been utilized to intercalate  (i) metals in the graphene/SiC(0001) interface,\cite{briggsNanoscale2019epitaxial,briggs2020atomically} as well as (ii) 2D insulating GaN layers embedded between graphene bilayer and the SiC(0001) surface.\cite{al2016two} In this section, we will investigate the energetic  stability, structural characterization, and electronic properties of h-BN ML on G/SiC(0001).} 

\begin{figure*}
    \includegraphics[width=16cm]{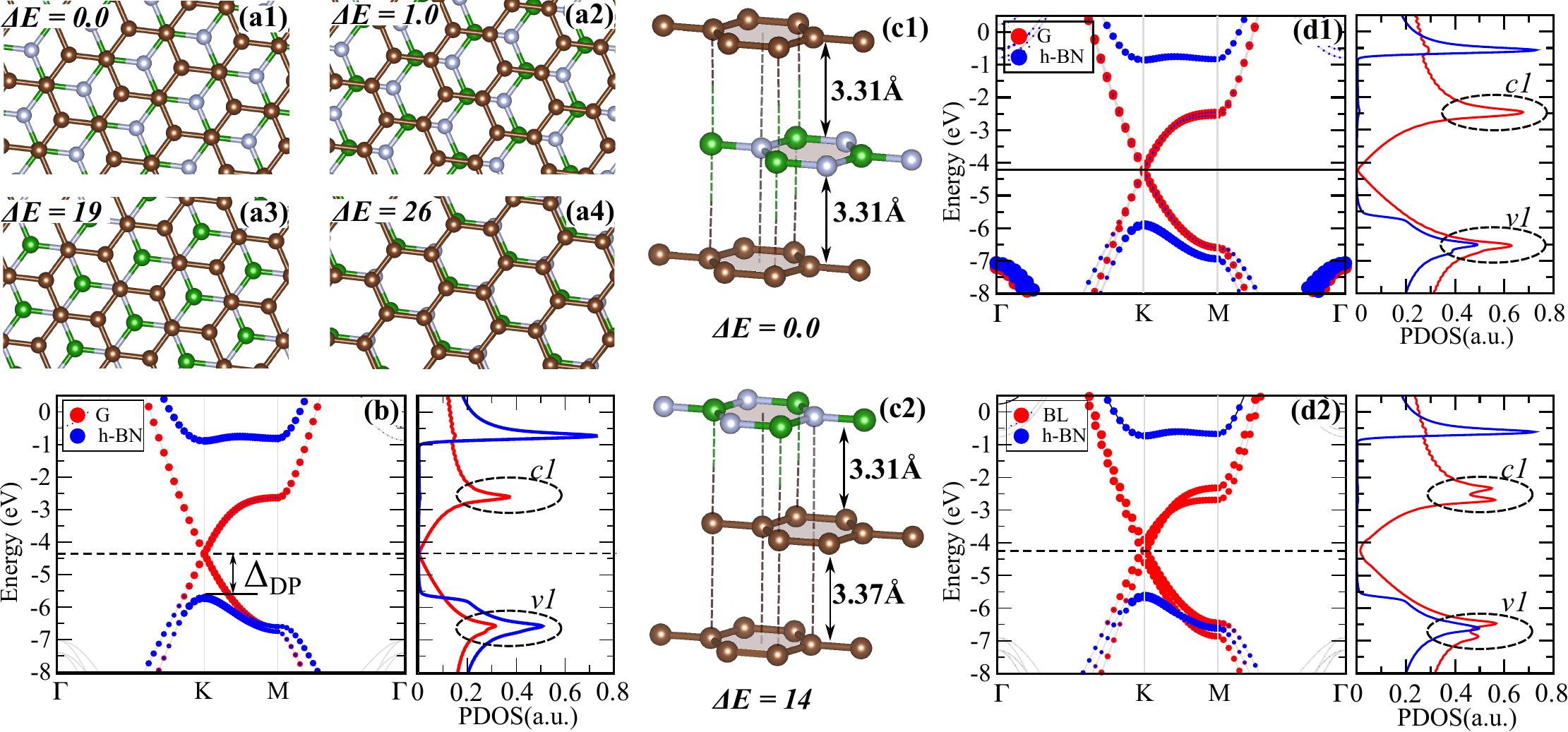}
    \caption{\label{models0} Structural models of G/h-BN interface,  Bernal stacking  with the boron atom aligned with the carbon atom (a1), boron-nitrogen bonding aligned with the carbon atom (a2), Bernal stacking with the nitrogen atom aligned with the carbon atom (a3),  eclipsed stacking (a4),   and the total energy differences [$\Delta E$ in meV/(1$\times$1)] with respect to the most stable configuration [(a1)]. (b) Electronic band structure, and projected density of states (PDOS) of the  G/h-BN bilayer (a1). Structural models of the trilayer systems, G/h-BN/G with (c1), h-BN/G/G (c2) with the boron atom aligned with the carbon atom, and  $\Delta E$ indicate the total energy difference  [in meV/(1$\times$1)] with respect to the most stable configuration [(c1)].  Electronic band structure, and  PDOS of the trilayer systems, G/h-BN/G (d1) and h-BN/G/G (d2) The zero energy is set at the vacuum level, and the carbon, boron, and nitrogen atoms are represented by brown, green and white spheres.}
\end{figure*}

\section{Computational details}

\subsection{Structural optimizations, total energies, and  electronic properties}

Our DFT calculations were performed within the Perdew-Burke-Ernzehof generalized gradient approximation \cite{PBE}, using the projector augmented wave (PAW) potentials,\cite{paw} as implemented in the Vienna Ab-initio Simulation Package(VASP) \cite{vasp1, vasp2}. The structural optimizations were done, including van der Waals corrections (vdW)\cite{tckatchenkoPRL2009} until the forces on each atom were smaller than 0.01 eV/\AA~ and total energies converged within a 1$\times$10$^{-6}$~eV criterion. The Kohn-Sham (KS)wave functions were expanded in a plane-wave basis set with an energy cutoff ($E_\text{cut}$) of 600 eV. The sampling of the Brillouin-zone (BZ) was performed by using the Monkhorst-Pack (MP)\cite{mp} scheme with k-point meshes  of 5$\times$5$\times$1 and 10$\times$10$\times$1 for structural optimization and electronic properties, respectively.

A supercell possessing a periodicity of $(2\sqrt{3}\times2\sqrt{3})R30^{\circ}$ is considered to simulate the interface that is formed by G/h-BN layers and SiC(0001) covered by a carbon buffer layer (SiC). The SiC(0001) surface was described by using a slab with three SiC  bilayers and a vacuum region of 16\,\AA. Furthermore, we employ a supercell rotation method, in which graphene and h-BN sheets are rotated by 6.58$^{\circ}$ (23.4$^{\circ}$) with respect to the direction of the armchair (zigzag) in relation to the substrate. By this method, the degree of mismatch between h-BN, graphene monolayer and bilayers is reduced to 0.4\%. This model has been successfully applied in graphene bilayers  and graphene/SiC interfaces\cite{Li2010,PhysRevB.82.121416,PADILHA2019603}. As in these systems, it is expected that the weak van der Waals interaction, primarily responsible for the bonding between these systems, such a rotation does not induce any relevant perturbations in the electronic structure of the interface due to the aforementioned rotation.

\subsection{XANES and CLS}

The carbon K-edge X-ray absorption near-edge structure (XANES) and C-$1s$ core-level shift (CLS)  were simulated by using the X-spectra package\cite{bunuauPRB2013,xas2,taillefumierPRB2002} implemented in the Quantum-ESPRESSO (QE) code\cite{espresso}. The KS orbitals were expanded in a plane-wave basis set with $E_\text{cut}$ of  48 and 192\,Ry for the single-particle wave-function and total charge density, respectively, and MP  10$\times$10$\times$1 k-points for the BZ sampling. For the carbon K-edge XANES, the initial state corresponds to a C-$1s$ orbital ($\ket{1s}$) originating from the absorbing atom without a core-hole. This orbital is extracted by using the Gauge-Including Projector Augmented-Wave (GIPAW) method\cite{gipaw} to calculate the dipolar cross-section,
\begin{equation}
\sigma(\omega) \propto	\omega \sum_{n} | \bra{n} \textbf{ê} \cdot \textbf{r} \ket{1s}|^{2} \delta (\varepsilon_{n} - \varepsilon_{1s} - \hbar\omega).
\end{equation}
The final state ($\ket{n}$) is obtained through the self-consistent solution of the KS equations for the whole system taking into account  the core-hole effects in the pseudopotential of the (X-ray) absorbing atom\cite{xas2}; $\textbf{ê}$ indicates the polarization vector of the incident X-ray, and $\varepsilon_n$ ($\varepsilon_{1s}$) is the energy of the final, $\ket{n}$ (initial, $\ket{1s}$) single-particle energy.

The CLS was obtained by using the so-called $\Delta$SCF (Difference in Self-Consistent Field) approach\,\cite{PhysRevB.74.045430,cls1}, where the core-level binding energy (BE) can be written as, 
%\begin{equation}
%    \Delta\text{BE}_{\Delta SCF}= E^{(n-1)} - E^{(n)}.
%\end{equation}
\begin{equation}
    \text{BE} = E^{(n-1)} - E^{(n)}.
    \label{BE}
\end{equation}
In this approach, $E^{(n-1)}$ and $E^{(n)}$ are the total energies of the ionized system with  $(n-1)$ electrons (a core-hole), and of the ground system with $n$ electrons, respectively. The former energy was obtained by using a pseudopotential generated based on the core-excited configuration of the atom \cite{PhysRevLett.71.2338}. 
Following the steps outlined in Refs.\,\cite{bolognesi2009investigation,bindingenergyBE2,castrovilli2018experimental}, we calculated $E^{(n)}$ using a regular pseudopotential (i.e., with no core-hole) and $E^{(n-1)}$ using a pseudopotential that included a core-hole in the investigated nonequivalent graphene carbon atoms. The same calculations were performed for a reference system, here a CH$_4$ molecule, resulting in BE$_0$ for the molecule's carbon atom.  The difference between BE and BE$_0$ results in the  C-1s core-level-shift, 
\begin{equation}  
\Delta\text{BE}=\text{BE}-\text{BE}_0.
\label{deltaBE}
\end{equation} 
 It is important to note that the molecule (reference system) is placed in the supercells in such a way that it does not interact with the SiC slab. Finally, the comparison with the experimental results can be done by adding the measured binding energy of the reference system ($\text{BE}_0^\text{exp}$), namely $\Delta\text{BE}+\text{BE}_0^\text{exp}$. For instance, using  the measured binding energy of C-$1s$ of the CH$_4$ molecule, $\text{BE}_0^\text{exp}$\,=\,283.5\,eV\cite{larciprete2001photochemistry}, we obtained a graphene's C-$1s$ core level BE of 284.07\,eV, which is in good agreement with experimental results for (quasi-free standing) graphene monolayer on SiC\cite{coletti2011large}.

\section{Results and Discussions}

\subsection{Free-standing bilayers and trilayers}

We start our investigation by looking at the energetically most stable stacking geometry of the free-standing G/h-BN BL. The interface pattern characterized by Bernal (AB) stacking with the boron atom aligned with the carbon atom [Fig.\,\ref{models0}(a1)], hereinafter referred to as G/h-BN, was found to be the most stable among the interfaces presented in Figs.\,\ref{models0}(a1)-(a4) \cite{giovannettiPRB2007substrate,fan2011tunable, kim2013synthesis}. At the equilibrium geometry, we find an interlayer distance of 3.32\,\AA. Our total charge density results reveal the absence of chemical bonds between the graphene and h-BN layers; the G/h-BN  binding strength is dictated by vdW long-range dispersive forces and no net charge  transfer takes place between  layers. In Fig.\,\ref{models0}(b) we present the orbital  projected electronic band structure of G/h-BN, characterized by (i) the Dirac bands of graphene lying within the energy gap of monolayer (ML) h-BN, with the Dirac point (DP) at 1.38\,eV above the valence band maximum (at the K-point) of h-BN, $\Delta_\text{DP}$\,=\,1.38\,eV, and (ii)  the emergence of a (small) energy gap of 35\,meV at the DP due to the graphene interaction with the h-BN layer\cite{pan2016modification,wang2017graphene}. The electronic density of states (DOS), projected on the $p_z$ orbitals,  indicates the formation of Van Hove singularities (VHSs) in G/h-BN\cite{indolese2018signatures}, labelled as $v1$ and $c1$ at $-2.2$ and $1.7$\,eV with respect to the Fermi level ($E_\text{F}$). It is worth noting that the electronic states at the M-point govern these VHSs, where $v1$ is the result of interlayer C-$p_z$+B-$p_z$ hybridization and $c1$ is solely due to C-$p_z$ orbitals. 

We next examined the structural model of TL systems composed of an h-BN ML encapsulated by graphene sheets (G/h-BN/G),  and on top of graphene bilayer (h-BN/G/G). We have considered the  Bernal stacking with the carbon atoms aligned with the boron atoms, and nitrogen atoms.  We found that G/h-BN/G  with the boron atoms aligned with the graphene carbon atom [Fig.\,\ref{models0}(c1)] as the energetically most stable configuration\cite{torres2022band}, followed by the h-BN/G/G  also with the boron-carbon alignment, Fig.\,\ref{models0}(c2). The other G/h-BN/G and h-BN/G/G configurations, with the carbon atom immediately above(or below) the nitrogen atom, are less stable by 16 and 34\,meV/(1$\times$1). At the equilibrium geometry, the interlayer  distances are practically identical to those found in G/h-BN. We have checked and confirmed these results by using other vdW approaches, Appendix-subsection\, A. 

The electronic band structure and the DOS projected on the $p_z$ orbitals of the G/h-BN/G TL, Fig.\,\ref{models0}(d1), resemble those of the G/h-BN BL. This is evidenced by the preservation of the linear dispersion at the K-point, followed by an energy gap of 8 meV at the DP\cite{torres2022band}, however, the energy position of the DP with respect to the VBM of h-BN increases to 1.64\,eV, $\Delta_\text{DP}$\,=\,1.38\,$\rightarrow$\,1.64\,eV. The projected DOS (PDOS) is characterized by the presence of VHSs, labelled as $v1$ and $c1$, at the same energy positions as those of G/h-BN. Both singularities are attributed to the electronic states at the M-point, with the former resulting from interlayer G/h-BN hybridizations and the latter from only the graphene $p_z$ states. We find a different picture in h-BN/G/G. In this case, the energy bands of the graphene layers are characteristics of a graphene bilayer. However,  the energy bands at the M-point are no longer degenerated since the stacking sequence (h-BN/G/G) is no longer mirror-symmetric, resulting in  doubled VHS in $v1$ and $c1$, as shown in Fig.\,\ref{models0}(d2)\cite{ramasubramaniam2011tunable}. Finally, it is worth noting that the work functions ($\Phi$) of the free-standing bilayer and trilayer systems (about 4.2\,eV), for both stacking geometries, are close to that of free-standing graphene\cite{mammadov2017work}, and consistent with the near absence of net charge transfers between the h-BN and graphene sheets.

%\textcolor{blue}{Note: In Ref.\,\cite{mammadov2017work} the author found $\Phi$ equal to 4.16 and 4.30\,eV, for graphene ML and BL on the SiC surface, and the Dirac point (DP) at 0.40 and 0.24\,eV below the Fermi level.} 

\subsection{h-BN/G BLs on buffer layer/SiC(0001)}

\subsubsection{Energetic and electronic properties}

Epitaxial graphene on SiC(0001) covered by a carbon buffer layer, here denoted as G/SiC [Fig.\,\ref{models0-ap}(a) (Appendix B)] has been extensively studied in the past few years. In consonance with the current literature, we find that (i) the graphene binding strength is determined by the vdW interaction, with a binding energy of 20\,meV/(1$\times$1), and (ii) there is a net charge transfer from SiC surface to the upper single-layer graphene, which results in a $n$-type doping of graphene at about $2.3\times 10^{13}e/\text{cm}^2$ \cite{mattauschPRL2007,varchonPRL2007,pankratov2012, sclauzeroPRB2012}. Such an electron doping results in a reduction of the graphene work function compared to that of single-layer free-standing graphene, namely $\Phi$\,=\,4.17\,$\rightarrow$\,3.69\,eV. 

%\textcolor{blue}{To be included, hydrogenated SiC surface.}

%\textcolor{red}{Upon saturation of the SiC surface dangling bonds with hydrogen atoms [Fig.\,\ref{models0-ap}(b) (Appendix)], (i) the graphene biding energy is nearly the same as that discussed above, whereas (ii) the SiC\,$\rightarrow$\,G net charge tranfer reduces to XXX$\times 10^{13}e/\text{cm}^2$. In this case,  work function of adsorbed graphene sheets is essentially the same as that of free-standing graphene.\,[$\leftarrow$\,Can be removed.]}

\begin{figure}
    \includegraphics[width=\columnwidth]{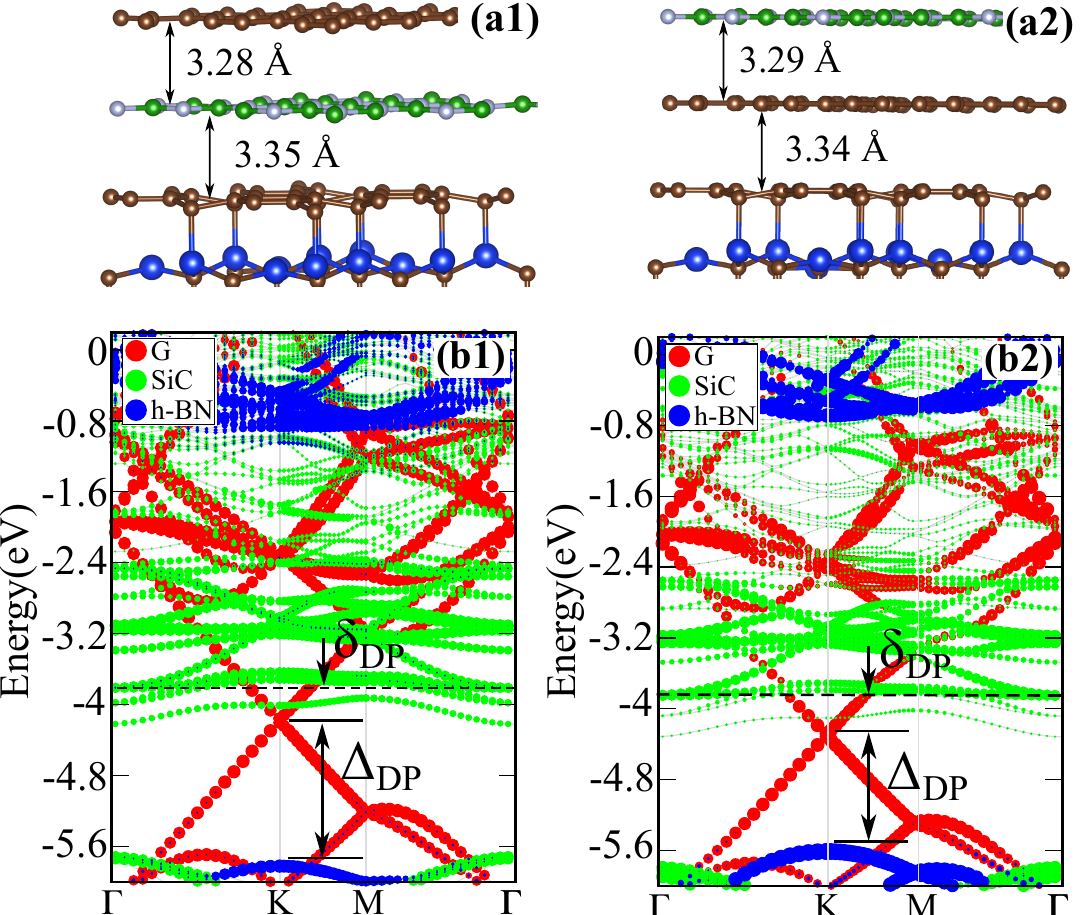}
    \caption{\label{models2} Structural models of h-BN ML (a1) on-top of the G/SiC(0001) surface [h-BN/G/SiC], and (a2) encapsulated between graphene single layer and the SiC surface, [G/h-BN/SiC]. Orbital projected electronic band structure of G/h-BN/SiC(0001) (b1), and h-BN/G/SiC (b2). The electronic density of states is proportional to the size of the filled circles, with  blue circles (red circles) corresponding to the projection on the boron and nitrogen (carbon) $p$-orbitals, and green circles to the $p$-orbitals of the surface Si dangling bonds. The zero energy was set at the vacuum level, and the black dotted line indicates the Fermi level.}
\end{figure}

Next, we consider heterostructures combining an h-BN/G BLs on buffer layer/SiC(0001). In Figs.\,\ref{models2}(a1) and (a2) we show the SiC(0001) surface covered by the carbon buffer layer and an h-BN ML below (G/h-BN/SiC) and above (h-BN/G/SiC) single-layer graphene. Our total energy results indicate that the G/h-BN/SiC is energetically more stable by 36\,meV/(1$\times$1) in comparison with  the latter stacking structure.  The interlayer binding strength is mediated mainly by vdW interactions, with an equilibrium distance of about 3.3\,\AA\, independent of the stacking order. 
%This is in accordance with our findings for the free-standing systems, Fig.\,\ref{models0}(c).

In both systems, the linear energy dispersion at the K-point has been preserved, Fig.\,\ref{models2}(b), characterized by a small energy gap at the K-point due to the lack of the A and B sublattice  symmetry\cite{zhouNatMat2007}. Moreover, it is noticeable that the energy positions of the h-BN's occupied states are downshifted in G/h-BN/SiC compared to those of h-BN/G/SiC, resulting in $\Delta_\text{DP}$\,=\,1.64 and 1.29\,eV, respectively, indicating that the h-BN states in G/h-BN/SiC are more tightly bonded than those in h-BN/G/SiC. Calculating the averaged orbital binding energy of the h-BN states  ($\bar\epsilon$) allows one to see this assertion in quantitative terms. Here $\bar\epsilon$ can be written as, 
\begin{equation}
\bar\epsilon = \frac{\int_{-\infty}^{E_\text{F}}\epsilon g(\epsilon)d\epsilon}{\int_{-\infty}^{E_\text{F}}g(\epsilon)d\epsilon},
\label{eq:band-energy}
\end{equation}
where $\epsilon$ is the single-particle energy level  and $g(\epsilon)$ is the respective PDOS of h-BN in G/h-BN/ and h-BN/G/SiC. We found that consistent with the energetic preference for the encapsulated h-BN, the electronic states of h-BN in G/h-BN/SiC are more tightly bonded than those in h-BN/G/SiC by 66\,meV, with the out-of-plane $p_z$ orbitals accounting for the major contribution (67\%).

\begin{figure}
    \includegraphics[width=\columnwidth]{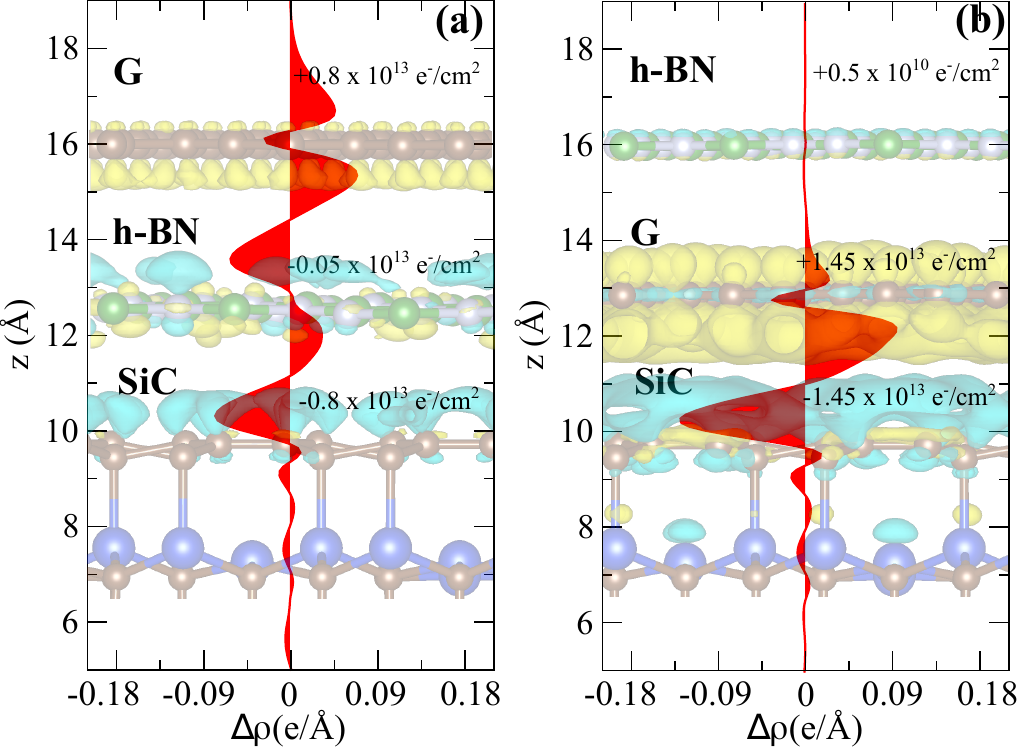}    \caption{\label{models3} Map of the electronic charge transfers [$\Delta\rho({\bf r})$], and the planar averaged $\Delta\rho$ [$\Delta\rho(z)$] of the bilayer/SiC systems, G/h-BN/SiC (a), h-BN/G/SiC (b). In the background, we show the electronic charge transfer isosurface for each stacking. Here the isosurface value is 2$\times$10$^{-4}$ e/\AA$^{3}$. Yellow (Blue) isosurfaces indicate $\Delta\rho>0$ ($<0$)}.
\end{figure}

The charge doping through the layered systems will depend on the G and h-BN stacking
orders. In Figs.\,\ref{models3}(a) and (b) we present the net charge transfers
($\Delta\rho$)  along  G/h-BN/ and h-BN/G/SiC\footnote{Here we have defined $\Delta\rho$
as $\rho_\text{G/h-BN/SiC}$-($\rho_\text{G/h-BN}$+$\rho_\text{SiC}$) in G/h-BN/SiC, and
 $\rho_\text{h-BN/G/SiC}$-$(\rho_\text{h-BN/G}$+$\rho_\text{SiC}$) in h-BN/G/SiC.}. The
charge transfers are ruled by the out-of-plane $p_z$ orbitals of the graphene adlayer,
h-BN, and the carbon  buffer layer. In G/h-BN/SiC, we found that the $n$-type doping of
graphene ($\Delta\rho>0$) reduces from  $\approx$~$2\times 10^{13}e/\text{cm}^2$ (in
G/SiC, not shown) to $0.8\times 10^{13}e/\text{cm}^2$, followed by (ii) an electrical
polarization of the h-BN layer with $\Delta\rho>0$ ($<0$) below (above) the sheet,
Fig.\,\ref{models3}(a). In this case, the intercalated h-BN sheet acts as a block layer to
the SiC\,$\rightarrow$\,G charge transfer. Meanwhile, we found that $n$-type doping of
graphene in h-BN/G/SiC is almost identical to that of G/SiC,  namely
$\Delta\rho$\,=\,$1.45\times 10^{13}e/\text{cm}^2$ in Fig.\,\ref{models3}(b), indicating
that the topmost h-BN sheet very slightly affects the charge transfer profile. Thus,
despite having identical interlayer equilibrium geometries, G/h-BN/ and h-BN/G/SiC exhibit
substantially distinct net charge transfers, which in its turn has an impact on the
electronic band structures\cite{mammadov2017work}. For instance,  as shown in
Figs.\,\ref{models2}(b1) and (b2), the DP lies at $\delta_\text{DP}$\,=\,0.28 and 0.32\,eV
below the Fermi level ($E_\text{F}$)  in G/h-BN/ and h-BN/G/SiC, respectively, which is
consistent with the lower doping rate in G/h-BN/SiC when compared with that of h-BN/G/SiC.
 By considering the vacuum level ($E_\text{vac}$) as the energy reference, we find the
graphene DP at 4.15 and 4.27\,eV below $E_\text{vac}$ in G/h-BN/ and h-BN/G/SiC. It is
possible to measure such difference in work function via Kelvin probe force microscopy
(KPFM)\cite{yu2009tuning,shtepliuk2017role,mammadov2017work}. Thus, this method can be
seen as a powerful tool for identifying the stacking order of large-area G/h-BN
heterostructures grown via physical- or chemical-vapor
deposition\cite{lopesProgCrystGrow2021synthesis}.

%%%%%%
In addition to the downshift  of $\bar\epsilon$, the energetic preference for the G/h-BN/SiC system can be attributed to the differences in the net charge transfer profiles between G/h-BN/ and h-BN/G/SiC  [Fig.\,\ref{models3}]. In the former, we have a negatively charged G layer neighboring a positively charged h-BN layer, while, in the latter, we have both G and h-BN negatively charged, resulting in a lower interlayer electrostatic energy in G/h-BN/SiC when compared with that of h-BN/G/SiC. Indeed, our total results show that  the classical electron-electron interaction, Hartree energy, brings the major contribution to the energetic preference of G/h-BN/SiC in comparison with the other stacking geometries.  Thus, we can infer that the net charge transfers through the G/h-BN stacked layers, upon its interaction with the SiC surface, also make an important contribution to the energetic stability of the h-BN encapsulated geometry, G/h-BN/SiC.

%\textcolor{red}{Is it possible to measure such a difference in the Work-function, for instance through Kelvin probe microscopy?\,\cite{yu2009tuning,shtepliuk2017role,mammadov2017work} If yes, we can say (or infer) that we may identify the stacking order based on in Kelvin probe measurements. }

\subsubsection{Structural characterization: XANES}
 
\begin{figure}[!htp]
    \includegraphics[width=\columnwidth]{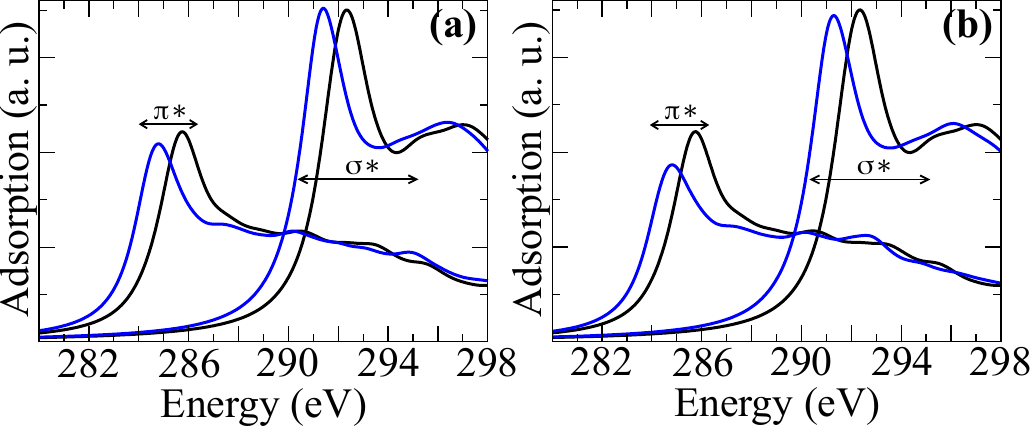}
    \caption{\label{xas1} Simulated C K-edge XANES of G/h-BN/SiC (a) and h-BN/G/SiC (b). The zero energy is  set to the calculated highest occupied state plus the experimental C-1$s$ binding energy.\cite{greczynski2020x}}
\end{figure}

In order to provide a more complete picture and connection between the stacking geometry and the electronic structure of the G/h-BN/ and h-BN/G/SiC systems, we performed a set of XANES simulations based on the first-principles DFT calculations.   Initially, we examined the K-edge absorption spectra of the graphene's carbon atoms, with the light polarization vector perpendicular ($\epsilon_\perp$) and parallel ($\epsilon_\parallel$) to the surface. Figure\,\ref{xas1} (solid-black lines) shows our simulated C K-edge spectra of pristine graphene, characterized by the  C-$1s$\,$\rightarrow$\,$\pi^\ast$,  and $\rightarrow$\,$\sigma^\ast$ transitions, at 285.8 and 292.4\,eV (for $\epsilon_\perp$ and $\epsilon_\parallel$). This results in a peak-to-peak difference [$\Delta E^\text{pp}=E(\sigma^\ast)-E(\pi^\ast)$] of 6.61\,eV, in good agreement with the experimental measurements\cite{schiros2012connecting,ouerghi2012large,lippitz2013plasma}.  In the sequence, we calculated the absorption spectra of the graphene carbon atoms in G/h-BN/ and h-BN/G/SiC [solid-blue lines in Figs.\,\ref{xas1}(a) and (b), respectively]. 
In agreement with the recent experimental XANES results of h-BN on  SiC(0001)\cite{sediriSciRep2015atomically}, we find that the  C-$1s$\,$\rightarrow$\,$\pi^\ast$,  and $\rightarrow$\,$\sigma^\ast$ absorption features of the graphene carbon  atoms have been preserved, as well as the energy difference $\Delta E^\text{pp}$, supporting the  vdW interactions between the G and h-BN layers\cite{sediriSciRep2015atomically}. When compared with  pristine graphene, we found that the absorption spectra in G/h-BN/ and h-BN/G/SiC are downshifted by about 1\,eV, where the  C-$1s$\,$\rightarrow$\,$\sigma^\ast$ absorption features have been preserved. In contrast, it is noticeable that the C-$1s$\,$\rightarrow$\,$\pi^\ast$ spectra show distinct features in G/h-BN/ and h-BN/G/SiC. In the former, because the graphene sheet lies on the h-BN/SiC surface, the absorption feature is almost  identical to that of the pristine system, whereas, in the latter, we find that the intensity of the absorption spectrum has been lowered upon the graphene's encapsulation by h-BN.  These findings suggest  that the stacking order of G and h-BN layers can be determined by the XANES spectra's ability to detect variations in the C-$2p_z$ hybridizations in G/h-BN/ and h-BN/G/SiC. 

% G/BN/SiC: pi-ratio=1.08
% BN/G/SiC: pi-ratio=1.19
% pristine graphene: pi - sigma = 6.61 eV 
%\textcolor{blue}{To be included, surface core level shifts (SCLSs) of C-$1s$, based on the final-state approximation,\cite{andersen1994surface} and compare with experimental results.\cite{coletti2011large,ouerghi2012large,sediriSciRep2015atomically}}

% grafeno, 284.07 eV.
%Mol + G/BN 	        283,50 	283,81 	283,77
%Mol + G/BN + Buffer 	283,50 	282,88 	282,88
%Mol +BN/G + Buffer 	283,50 	283,05 	283,04

\subsection{G/h-BN/G TLs on buffer layer/SiC(0001)}

\subsubsection{Energetic and electronic properties}

\begin{figure}[!htp]
    \includegraphics[width=\columnwidth]{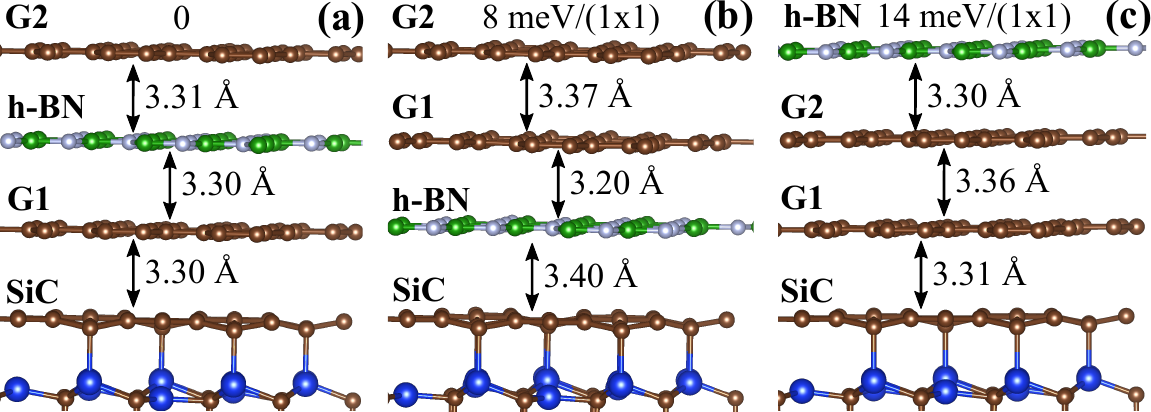}
    \caption{Schematic models for three possible stackings of the trilayers G2, G1, and h-BN layers:  (a) G2/BN/G1, (b) G2/G1/BN, and (c) BN/G2/G1. We also have shown the interlayer distances (in angstrom) in the figure. In addition, at the top, the total energy differences between the most stable stacking (a), which we have defined as zero energy.\label{models4} }
\end{figure}

In this last subsection, we examine the structural/energetic, and electronic properties of the graphene and h-BN TL systems on buffer layer-covered SiC(0001) (trilayer/SiC), Fig.\,\ref{models4}. We found that the energetic preference for the h-BN intercalated (encapsulated) between graphene layers G2/h-BN/G1/SiC [Fig.\,\ref{models4}(a)] has been preserved, while the h-BN on the topmost surface layer, h-BN/G2/G1/SiC [Fig.\,\ref{models4}(c)] is the energetically less favorable configuration by 14\,meV/(1$\times$1). At the equilibrium geometry,  the interlayer distances  in G2/h-BN/G1/SiC are the same when compared with those of the free-standing G/h-BN/G heterostructure, 3.31\,\AA\ in Fig.\,\ref{models0}(c1).

\begin{table}
\caption{Graphene's C-$1s$ core-level shift ($\Delta\text{BE}$ in eV) of G/h-BN and h-BN/G on SiC with respect to the one of free-standing graphene.}
\begin{ruledtabular}
\begin{tabular}{ccc}
 %& \multicolumn{2}{c}{bilayer/SiC} & \multicolumn{3}{c}{trilayer/SiC} \\ \hline
                 & G/h-BN/   &  h-BN/G/  \\
                 \hline
 $\Delta\text{BE}$  & $-1.19$   &  $ -1.02$ \\ 
\end{tabular}
\end{ruledtabular}
\label{table:cls1} 
\end{table}
\begin{table}
\caption{Graphene's C-$1s$ core-level shift  ($\Delta\text{BE}$ in eV) of G2/h-BN/G1, G2/G1/h-BN, and h-BN/G2/G1 on SiC with respect to the one of free-standing graphene.}
\begin{ruledtabular}
\begin{tabular}{c|c|c|c}
                    & G2/h-BN/G1/ & G2/G1/h-BN/ & h-BN/G2/G1/ \\
                    \hline
$\Delta\text{BE}$\,(G2/G1)   &  $-1.32/-1.49$ & $-1.29/-1.36$ & $-1.39/-1.21$  \\
\end{tabular}
\end{ruledtabular}
\label{table:cls2} 
\end{table}

\begin{figure}[htp]
    \includegraphics[width=\columnwidth]{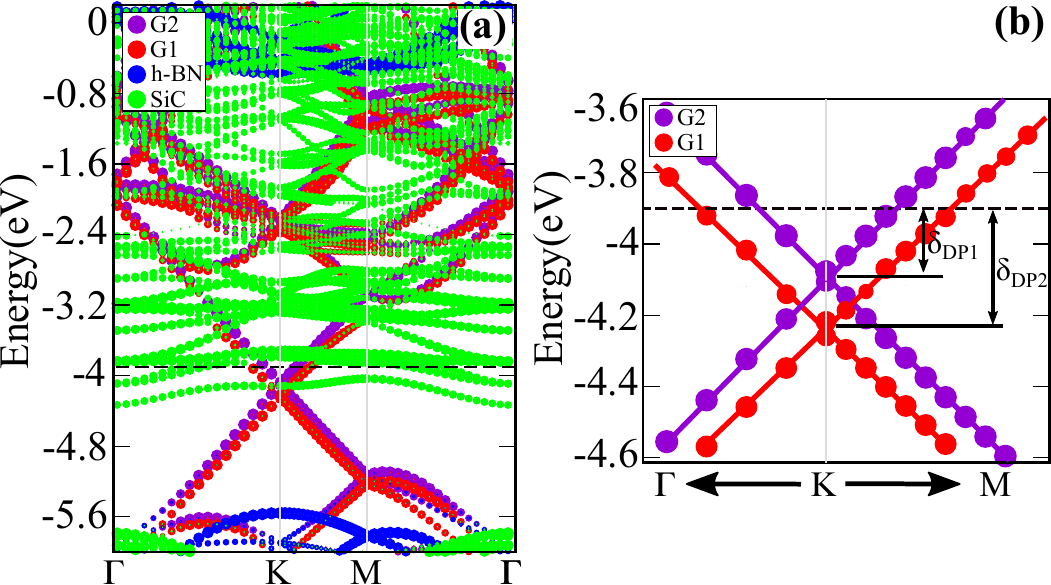}
    \caption{(a)  Projected states band structure of the single graphene layers G1-$p$ states (red sphere) and G2-$p$ states (purple sphere), BN-$p$ states (blue ball) and Buffer [Si-$p$, C-$p$ and H-$p$ states] (green sphere) em G2/BN/G1 on the buffer. (b) Dirac cones are detailed around the Fermi level. The zero energy was set at the vacuum level, and the black dotted line indicates the Fermi level. \label{bands4} }
\end{figure}

Figure\,\ref{bands4}(a) depicts the orbital projected electronic band structure of G2/h-BN/G1/SiC. It is possible to observe that (i) the  linear energy dispersion of graphene has been preserved, resulting in (ii)   two non-degenerated Dirac energy bands. In contrast, the electronic band structures of G2/G1/h-BN/ and h-BN/G2/G1/SiC are ruled by the formation of graphene bilayer systems, G2/G1, characterized by the  emergence of parabolic bands near the Fermi level, as depicted in Figs.\,\ref{fig:bands5}(a) and (b) Appendix C. 

\begin{figure*}[htp]
    \includegraphics[width=15cm]{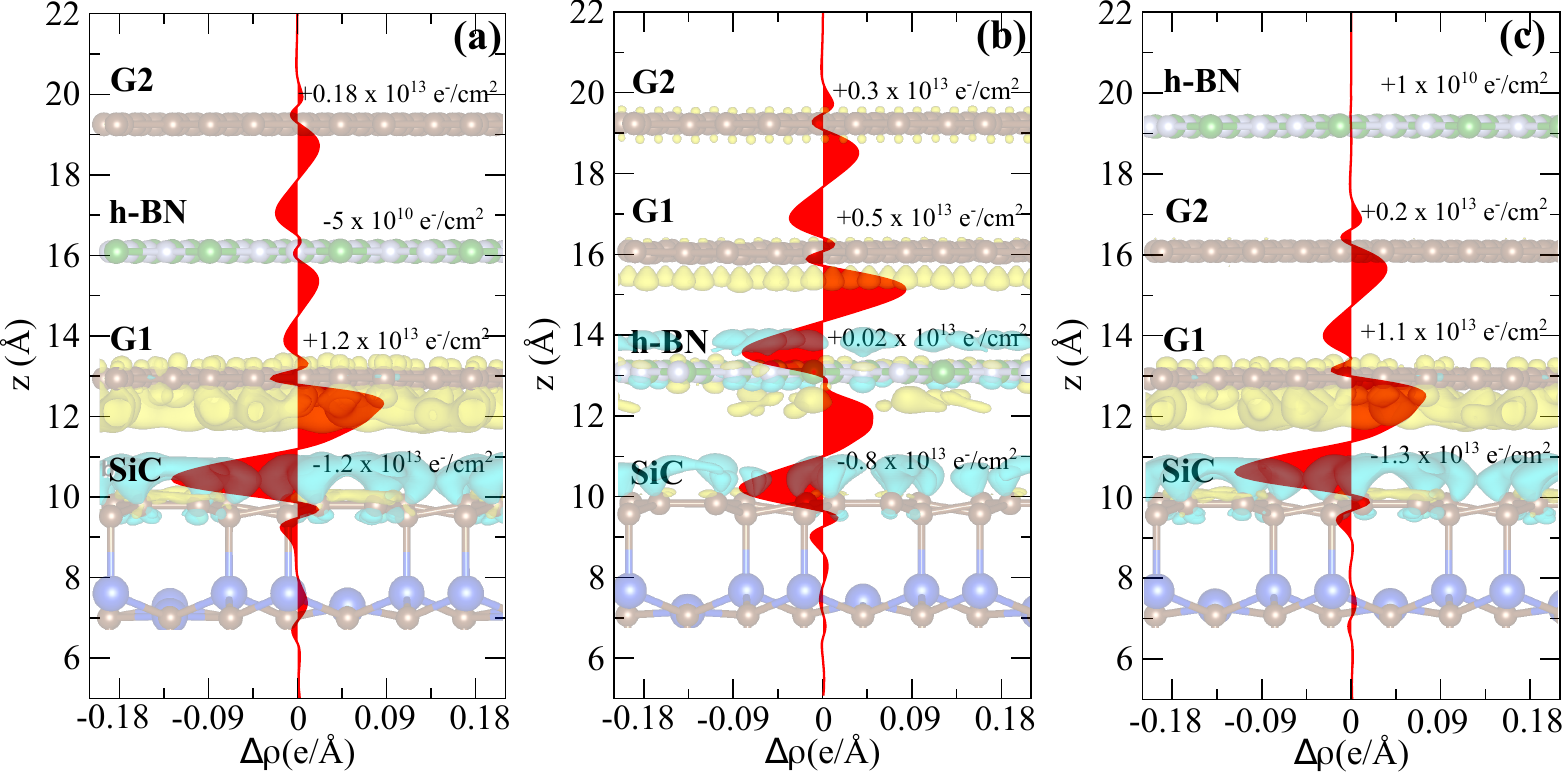}
    \caption{Map of the electronic charge transfers [$\Delta\rho({\bf r})$], and the planar averaged $\Delta\rho$ [$\Delta\rho(z)$] of the trilayer/SiC systems, G2/h-BN/G1/SiC (a), G2/G1/h-BN/SiC (b), and h-BN/G2/G1/SiC (c). Yellow (Blue) isosurfaces indicate $\Delta\rho>0$ ($<0$). The isosurface value is 2$\times$10$^{-4}$ e/\AA$^{3}$. \label{fig:deltarhotrilayer} }
\end{figure*}

In G2/h-BN/G1/SiC, the Dirac points DP1 and DP2  lie below the Fermi level [Fig.\,\ref{bands4}(b)]   with $\delta_\text{DP1}$ and $\delta_\text{DP2}$ of $-0.18$ and $-0.34$\,eV, respectively, dictated by the charge transfers from the SiC surface to the G2/h-BN/G1 trilayer, see Fig.\,\ref{fig:deltarhotrilayer}(a). Such charge transfers are mostly localized at the G1/SiC interface, resulting in  a $n$-type doping of G1 by $\Delta\rho_\text{G1}$\,=\,$1.2\times 10^{13}e/\text{cm}^2$, which is smaller than the doping level obtained in G/SiC, while the positive net charge density ($\Delta\rho_\text{SiC}$\,=\,$-1.2\times 10^{13}e/\text{cm}^2$) is predominantly localized on the topmost layer of SiC. The presence of the intercalated h-BN layer decreases the electron doping of G2 to  $\Delta\rho_\text{G2}$\,=\,$0.18\times 10^{13}e/\text{cm}^2$, while h-BN exhibits a very slight hole doping of $5\times 10^{10}e/\text{cm}^2$.

%%%%%%%%%%%%%%%%%%
%\textcolor{red}{Here, comparing our $\Delta\rho$ result  in G1/h-BN/G2/SiC and h-BN/G/SiC, 1.2 and $2.3\times 10^{13}e/\text{cm}^2$, respectively,  it is not clear why such a reduction in the trilayer systems; seems that somehow the graphene bilayer plays a role, but how?} 
%%%%%%%%%%%%%%%%%%

The amount of SiC\,$\rightarrow$\,trilayer charge transfers are roughly the same regardless of stacking order, Fig.\,\ref{fig:deltarhotrilayer}. As shown in Figs.\,\ref{fig:deltarhotrilayer}(b) and (c),  in the G2/G1/h-BN and h-BN/G2/G1/SiC systems, the negatively charged layers G1 and G2 are stacked in sequence, suggesting in higher electrostatic energy when compared to G2/h-BN/G1/SiC, where the negatively charged layers G1 and G2 are separated by the positively charged h-BN. Indeed, similar to what we verified in the bilayer/SiC systems, our total energy results reveal that   the energetic preference of G2/h-BN/G1/SiC is dictated by the  Hartree energy component. It is worth noting that in the free-standing trilayer systems [Fig.\,\ref{models0}(c)] the energetic preference for G/h-BN/G is ruled by the electrostatic ion-ion Coulomb energy (Ewald term). In addition, further calculation of the h-BN's orbital binding energy [eq.\,\ref{eq:band-energy}] support the energetic preference for the encapsulated geometry in trilayer/SiC.

%Thus, suggesting that the energetic preference for the stacking order of graphene and h-BN layers in bilayer/ and trilayer/SiC is ruled by the reduction of the interlayer electrostatic energy, which is in its turn ruled by the electron/hole doping profile.}

%The orbital projected electronic band structure of G2/h-BN/G1/SiC [Fig.\,\ref{bands4}(b)] shows that the  graphene layers, G1 and G2, give rise to two non-degenerated Dirac energy bands. Because of the electron doping of G1 and G2, the  Dirac points DP1 and DP2 are found below the Fermi (vacuum) level  with $\Delta_\text{DP1}$ and $\Delta_\text{DP2}$ of 0.18 and 0.34\,eV (XXX and XXX\,eV).\cite{mammadov2017work}  

\subsubsection{Structural characterization: CLS}

Recent works, based  X-ray photoelectron spectroscopy (XPS) experiments, have indicated the formation of sharp h-BN/G interfaces on the SiC substrate, with the absence of B clusters or the formation of chemical bonds between  graphene and h-BN\cite{sediriSciRep2015atomically,heilmann2DMat2018defect}. Motivated by the recent successful combination of experimental results and first-principles DFT simulations leading to an atomistic understanding of XPS spectra\cite{ugolotti2017chemisorption,castrovilli2018experimental,golze2022accurate},here we start an investigation of the graphene's C-$1s$ core level binding energy (BE) of the bilayer/ and trilayer/SiC systems. We focused on the dependence of the core level binding energy, i.e., the core-level shift (CLS), on G/h-BN stacking geometry and net charge transfers in bilayer/ and trilayer/SiC, Figs.\,\ref{models3} and \ref{fig:deltarhotrilayer}.

%Based on the calculation procedure described in Sec.\,II, we obtained a graphene's C-$1s$ core level BE of 284.07\,eV\cite{BE}, which is in good agreement with experimental results for (quasi-free standing) graphene monolayer on SiC.\cite{coletti2011large} 
%%%%%%%%%%%%
%\textcolor{red}{[It is worth considering the BE of G/SiC, with no h-BN, as a reference for the core-level shifts discussed below. To be calculated.]}
%%%%%%%%%%%%
We found that the graphene's C-$1s$ core level BE reduces upon the formation of bilayer/ and trilayer/SiC systems. For instance, in G/h-BN/SiC, the SiC\,$\rightarrow$\,G/h-BN net charge transfers results in a reduction of the C-$1s$ core level binding energy of 1.19\,eV with respect to the one of free-standing graphene,  $\Delta\text{BE}$\,=\,$-1.19$\,eV.
%%%%%%%%%%%
%\textcolor{red}{[If we consider the C-$1s$'s BE of G/SiC, instead fo free-standing graphene (284.07\,eV), we will find lower values of $\Delta\text{BE}$.]}
%%%%%%%%%%%
By changing the stacking sequence, h-BN/G/SiC, we found $\Delta\text{BE}$\,=\,$-1.02$\,eV, thus indicating a stacking dependence of 0.17\,eV in the graphene's C-$1s$ core-level spectrum.  Our results of  $\Delta\text{BE}$, summarized in Tables\,\ref{table:cls1} and \ref{table:cls2}, reveal the manifestation of two main contributions to the core-level shifts, namely (i) net charge transfer (electron doping, $\Delta\rho$), and (ii) the presence of external electric field (EEF)\cite{bagus1999mechanisms}. For instance, based on the electron doping level [(i)] of the graphene layer in h-BN/G/SiC [Fig.\,\ref{models3}(b)], it is expected a larger reduction of the graphene's C-$1s$ core-level BE compared with that in G/h-BN/SiC. However, this is not what we found, which can be attributed to the emergence of an EEF [(ii)] due to the positively charged layer on the SiC surface [Fig.\,\ref{models3}(a)], which increases the BE of graphene C-$1s$, and thus leading to a smaller BE reduction in h-BN/G/SiC. Similarly in the trilayer/SiC system, we find that both, the electron doping [(i)], and the emergence of an EEF [(ii)], due to the trilayer/SiC interface dipole, rule the C-$1s$ core-level shifts, $\Delta\text{BE}$. Further discussions of the graphene C-$1s$ core-level shift are presented in Appendix D. It is worth noting that, despite the small energy differences ($\sim$\,$0.1-0.2$\,eV), our results of graphene's C-$1s$ core-level BEs reveal a relationship between the G/h-BN stacking order and the respective CLSs.

\section{Summary and Conclusions}

By means of first-principles DFT calculations, we have performed a theoretical study of G/h-BN van der Waals heterostructures on SiC(0001) covered by a carbon buffer layer. Our total energy results not only show the encapsulation of h-BN underneath epitaxial single-layer graphene in bilayer/ and trilayer/SiC systems (as inferred from recent experimental observations), but also reveal that the energetic preference for the h-BN encapsulation is ruled by a reduction of the  electron-electron classical electrostatic  (Hartree) energy, triggered by the electron doping of graphene.  We found that the graphene layers become $n$-type doped, where the doping profile through the vdW layers is determined by the stacking geometry. Further electronic band structure calculations reveal that the linear energy band dispersion of graphene has been preserved in G/h-BN/ and G/h-BN/G/SiC systems, with the Dirac point lying below the Fermi level. Finally, we found key spectroscopic fingerprints of bilayer and trilayer SiC using K-edge X-ray absorption near edge spectroscopy (XANES) and C-$1s$ core-level-shift (CLS) simulations, which may aid in the experimental characterization of these G/h-BN heterostructures on SiC.

\begin{acknowledgments}

The authors acknowledge financial support from the Brazilian agencies CNPq, CAPES, FAPEMIG, and INCT-Nanomateriais de Carbono, and the CENAPAD-SP and Laborat{\'o}rio Nacional de Computa{\c{c}}{\~a}o Cient{\'i}fica (LNCC-SCAFMat2) for computer time.

\end{acknowledgments}

%\section{Appendix}
\appendix

\section{Energetic stability of trilayers}

%It's known that the vdW-potential plays a crucial role in determining the structural and electronic properties (reference here). Therefore, we calculated the energetic stability of G/BN/G and  BN/G/G pristine stacking by considering, beyond the TS-vdW, the other three vdW potentials already implemented in the VASP and QE codes: optB88-vdW, optPBE-vdW and DF-vdW (see Table XX). We have admitted the Bernal stacking in which the carbon atoms aligned with the boron and nitrogen atoms. The results indicate that in any vdW-potential, the G/BN/G with boron and carbons aligned (it means the nitrogen atoms on the hole of graphene's hexagons) is ever the most stable stacking}.  

We have performed additional calculations of the energetic stability of the trilayer systems, G/h-BN/G and h-BN/G/G, using other approaches to describe the long-range vdW interactions, namely vdW-Opt88, -OptPBE,  -DF, and -TS. Our results, summarized in Table\,\ref{table:BL}, indeed confirm the energetic preference of the former staking geometry, with the graphene carbon atoms aligned with the B atoms of h-BN.

\begin{table}[!htp]
\caption{Total energies differences  (in meV/1$\times$1), of  the G/h-BN/G and h-BN/G/G stacking geometries with respect to the energetically most stable configuration [G/h-BN/G with the graphene's C atom aligned with the h-BN's B atom], by using different approaches to describe the vdW interactions.}

\begin{ruledtabular}
\begin{tabular}{lcccc}
 & \multicolumn{2}{c}{G/h-BN/G} & \multicolumn{2}{c}{h-BN/G/G} \\ 
\cline{2-3} \cline{4-5}
         &   C aling. B    & C aling. N     &     C aling. B & C aling. N   \\ 
OptB88   &    0    &    6     &     5  & 22   \\
OptPBE   &    0    &    7     &     3  & 15           \\
DF &    0    &    10     &     1 & 9           \\
TS &    0    &    16     &     8 & 32           \\
\end{tabular}
\end{ruledtabular}
\label{table:BL} 
\end{table}
 
 \section{G/SiC}
\begin{figure}[!htp]
    \includegraphics[width=\columnwidth]{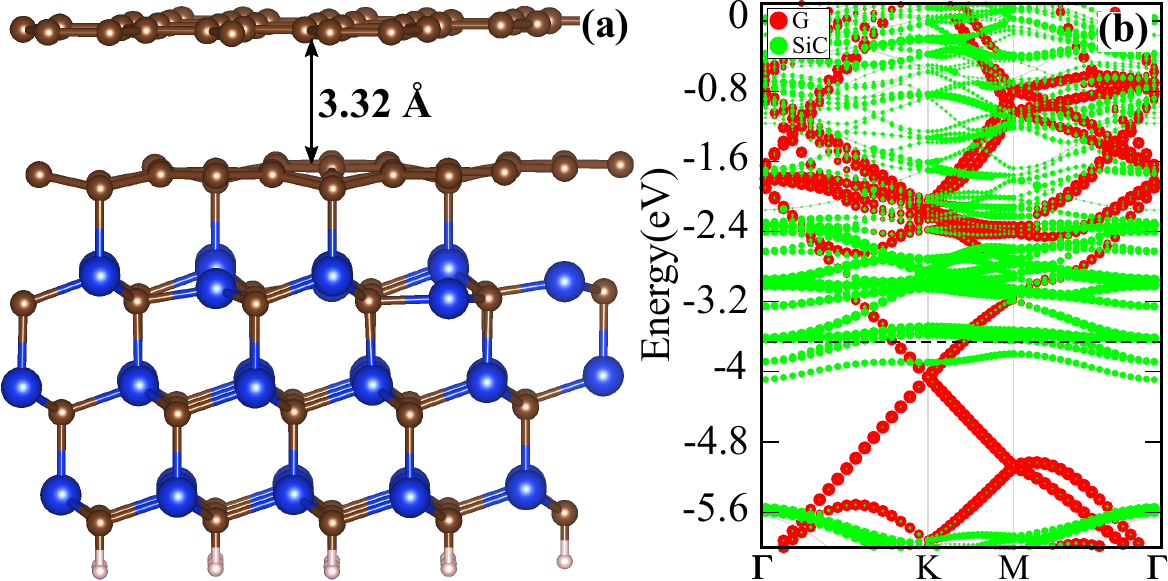}
    \caption{\label{models0-ap} Structural model (a) and the electronic band structure (b) of graphene on the SiC(0001) surface covered by a carbon buffer layer, G/SiC.}
\end{figure}

 Figures\,\ref{models0-ap}(a) and (b) show the structural model of G/SiC and electronic band structure of graphene on the SiC(0001) surface covered by a carbon buffer  layer, G/SiC, respectively.

\section{G2/G1/h-BN/ and h-BN/G2/G1/SiC}
\begin{figure}[htp]
    \includegraphics[width=\columnwidth]{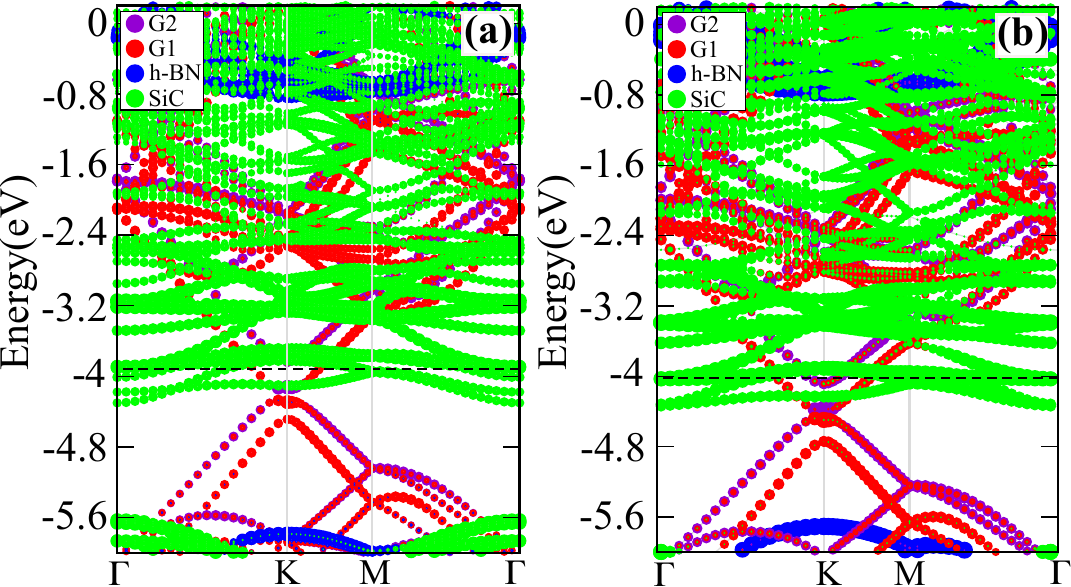}
    \caption{Orbital projected electronic band structures of G2/G1/h-BN/SiC (a), and h-BN/G2/G1/SiC (b). \label{fig:bands5} }
\end{figure}

In Figs.\,\ref{fig:bands5}(a) and (b) we present the orbital projected electronic band structures of the G2/G1/h-BN/ and h-BN/G2/G1/SiC systems, both characterized by the formation of massive bands below the Fermi level.

\section{\label{cls-ap}Core-level shifts of intermediate structures}
\begin{figure}[!htp]
    \includegraphics[width=\columnwidth]{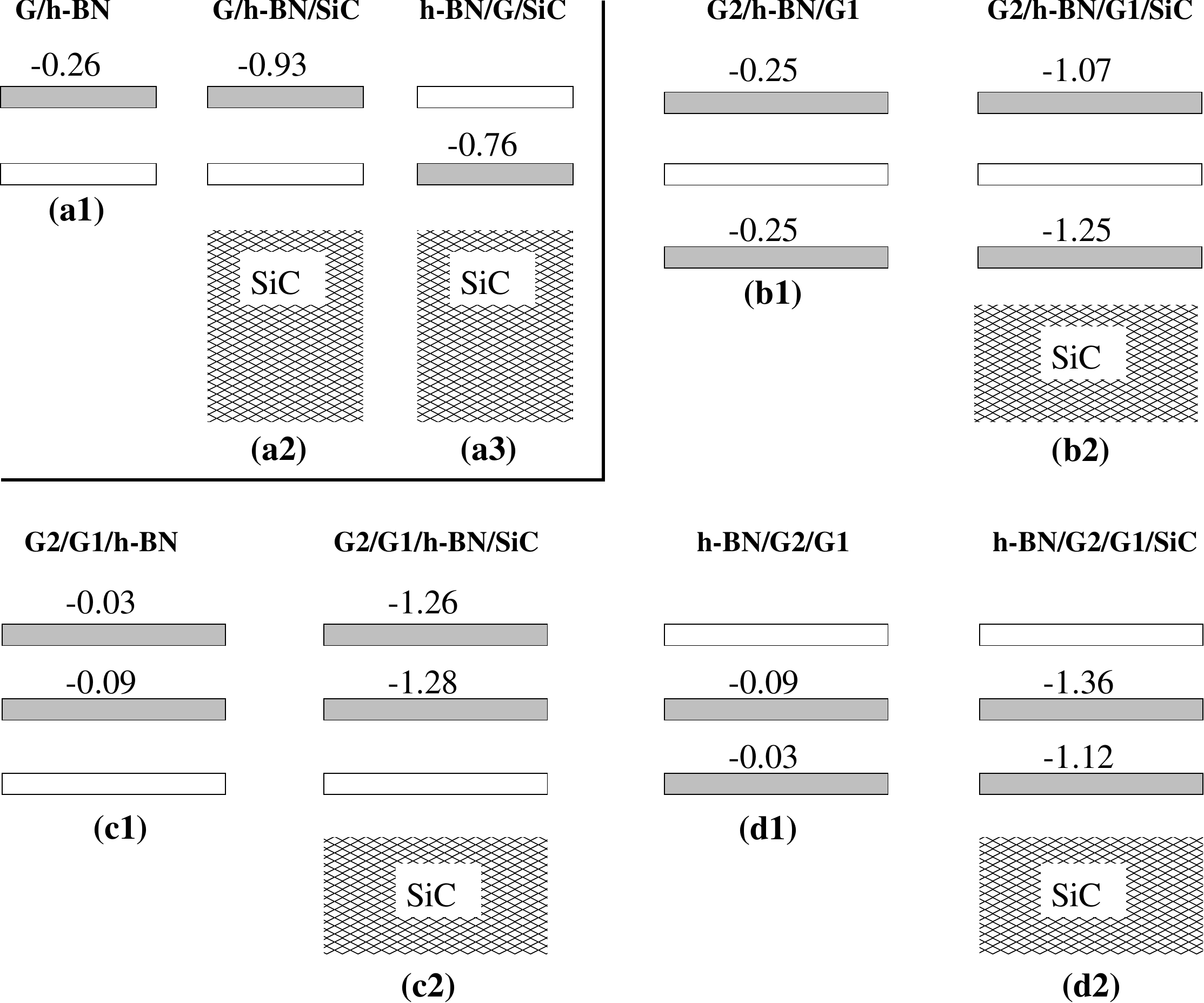}
    \caption{\label{cls}Schematic diagram of the graphene's C-$1s$ core-level shifts in the bilayer/SiC (a), and trilayer/SiC (b)-(d). (a1) $\Delta\text{BE}$ of G/h-BN with respect to the graphene monolayer, (a2) and (a3) $\Delta\text{BE}$ of G/h-BN bilyer upon its interaction with the SiC(0001) surface covered by a carbon buffer layer, SiC. (b1) $\Delta\text{BE}$ of G2/h-BN/G1 with respect to the graphene monolayer, (b2) $\Delta\text{BE}$ of G2/h-BN/G1 trilayer upon its interaction with the SiC surface. (c1) and (d1) $\Delta\text{BE}$ of G2/G1/h-BN and h-BN/G2/G1 with respect to the graphene monolayer, (c2) and (d2) $\Delta\text{BE}$ of G2/G1/h-BN and h-BN/G2/G1 trilayers upon their interaction with the SiC surface.}
\end{figure}
To better comprehend the graphene's C-$1s$ core-level shift in bilayer/ and trilayer/SiC, we calculate the BEs by  considering  (hypothetical) intermediate structures. For instance, in Fig.\,\ref{cls}(a) the hypothetical structure is represented by a free-standing G/h-BN bilayer fixed at the equilibrium geometry of the G/h-BN/SiC  final system. We find that BE reduces by 0.26\,eV, when compared  to one of the free-standing graphene, $\Delta\text{BE}$\,=\,$-0.26$\,eV; which can be attributed to the orbital overlap at the G/h-BN interface region.\cite{fan2011tunable} As expected, we found the same result of $\Delta\text{BE}$ when we consider the equilibrium geometry of h-BN/G/SiC. In contrast, the interaction of the G/h-BN bilayer with the SiC surface (final system) results in the stacking dependent values of $\Delta\text{BE}$, viz.: $\Delta\text{BE:}\,\text{G/h-BN}$\,$\xrightarrow{-0.93\,\text{eV}}$\,$\text{G/h-BN/SiC}$ and $\Delta\text{BE:}\,\text{h-BN/G}$\,$\xrightarrow{-0.76\,\text{eV}}$\,$\text{h-BN/G/SiC}$, as schematically shown in  Figs.\,\ref{cls}(a2) and (a3). Here, we can infer that such dependence is dictated by a competition between the electron doping of graphene and the electric field due to the formation of a positively charged layer on the SiC surface, where the former (latter) reduces (increases) the BE of graphene's C-1s core-level. A similar picture has been verified in the trilayer/SiC systems, as depicted in Figs.\,\ref{cls}(b)-(d).

\bibliography{main}

\providecommand{\noopsort}[1]{}\providecommand{\singleletter}[1]{#1}%
\begin{thebibliography}{61}%
\makeatletter
\providecommand \@ifxundefined [1]{%
 \@ifx{#1\undefined}
}%
\providecommand \@ifnum [1]{%
 \ifnum #1\expandafter \@firstoftwo
 \else \expandafter \@secondoftwo
 \fi
}%
\providecommand \@ifx [1]{%
 \ifx #1\expandafter \@firstoftwo
 \else \expandafter \@secondoftwo
 \fi
}%
\providecommand \natexlab [1]{#1}%
\providecommand \enquote  [1]{``#1''}%
\providecommand \bibnamefont  [1]{#1}%
\providecommand \bibfnamefont [1]{#1}%
\providecommand \citenamefont [1]{#1}%
\providecommand \href@noop [0]{\@secondoftwo}%
\providecommand \href [0]{\begingroup \@sanitize@url \@href}%
\providecommand \@href[1]{\@@startlink{#1}\@@href}%
\providecommand \@@href[1]{\endgroup#1\@@endlink}%
\providecommand \@sanitize@url [0]{\catcode `\\12\catcode `\$12\catcode
  `\&12\catcode `\#12\catcode `\^12\catcode `\_12\catcode `\%12\relax}%
\providecommand \@@startlink[1]{}%
\providecommand \@@endlink[0]{}%
\providecommand \url  [0]{\begingroup\@sanitize@url \@url }%
\providecommand \@url [1]{\endgroup\@href {#1}{\urlprefix }}%
\providecommand \urlprefix  [0]{URL }%
\providecommand \Eprint [0]{\href }%
\providecommand \doibase [0]{http://dx.doi.org/}%
\providecommand \selectlanguage [0]{\@gobble}%
\providecommand \bibinfo  [0]{\@secondoftwo}%
\providecommand \bibfield  [0]{\@secondoftwo}%
\providecommand \translation [1]{[#1]}%
\providecommand \BibitemOpen [0]{}%
\providecommand \bibitemStop [0]{}%
\providecommand \bibitemNoStop [0]{.\EOS\space}%
\providecommand \EOS [0]{\spacefactor3000\relax}%
\providecommand \BibitemShut  [1]{\csname bibitem#1\endcsname}%
\let\auto@bib@innerbib\@empty
%</preamble>
\bibitem [{\citenamefont {Geim}\ and\ \citenamefont
  {Grigorieva}(2013)}]{geim2013van}%
  \BibitemOpen
  \bibfield  {author} {\bibinfo {author} {\bibfnamefont {Andre~K}\ \bibnamefont
  {Geim}}\ and\ \bibinfo {author} {\bibfnamefont {Irina~V}\ \bibnamefont
  {Grigorieva}},\ }\bibfield  {title} {\enquote {\bibinfo {title} {Van der
  waals heterostructures},}\ }\href@noop {} {\bibfield  {journal} {\bibinfo
  {journal} {Nature}\ }\textbf {\bibinfo {volume} {499}},\ \bibinfo {pages}
  {419--425} (\bibinfo {year} {2013})}\BibitemShut {NoStop}%
\bibitem [{\citenamefont {Massicotte}\ \emph {et~al.}(2016)\citenamefont
  {Massicotte}, \citenamefont {Schmidt}, \citenamefont {Vialla}, \citenamefont
  {Watanabe}, \citenamefont {Taniguchi}, \citenamefont {Tielrooij},\ and\
  \citenamefont {Koppens}}]{massicotte2016photo}%
  \BibitemOpen
  \bibfield  {author} {\bibinfo {author} {\bibfnamefont {Mathieu}\ \bibnamefont
  {Massicotte}}, \bibinfo {author} {\bibfnamefont {Peter}\ \bibnamefont
  {Schmidt}}, \bibinfo {author} {\bibfnamefont {Fabien}\ \bibnamefont
  {Vialla}}, \bibinfo {author} {\bibfnamefont {Kenji}\ \bibnamefont
  {Watanabe}}, \bibinfo {author} {\bibfnamefont {Takashi}\ \bibnamefont
  {Taniguchi}}, \bibinfo {author} {\bibfnamefont {Klaas-Jan}\ \bibnamefont
  {Tielrooij}}, \ and\ \bibinfo {author} {\bibfnamefont {Frank~HL}\
  \bibnamefont {Koppens}},\ }\bibfield  {title} {\enquote {\bibinfo {title}
  {Photo-thermionic effect in vertical graphene heterostructures},}\
  }\href@noop {} {\bibfield  {journal} {\bibinfo  {journal} {Nature
  communications}\ }\textbf {\bibinfo {volume} {7}},\ \bibinfo {pages} {12174}
  (\bibinfo {year} {2016})}\BibitemShut {NoStop}%
\bibitem [{\citenamefont {Haigh}\ \emph {et~al.}(2012)\citenamefont {Haigh},
  \citenamefont {Gholinia}, \citenamefont {Jalil}, \citenamefont {Romani},
  \citenamefont {Britnell}, \citenamefont {Elias}, \citenamefont {Novoselov},
  \citenamefont {Ponomarenko}, \citenamefont {Geim},\ and\ \citenamefont
  {Gorbachev}}]{haigh2012cross}%
  \BibitemOpen
  \bibfield  {author} {\bibinfo {author} {\bibfnamefont {Sarah~J}\ \bibnamefont
  {Haigh}}, \bibinfo {author} {\bibfnamefont {Ali}\ \bibnamefont {Gholinia}},
  \bibinfo {author} {\bibfnamefont {Rashid}\ \bibnamefont {Jalil}}, \bibinfo
  {author} {\bibfnamefont {Simon}\ \bibnamefont {Romani}}, \bibinfo {author}
  {\bibfnamefont {Liam}\ \bibnamefont {Britnell}}, \bibinfo {author}
  {\bibfnamefont {Daniel~C}\ \bibnamefont {Elias}}, \bibinfo {author}
  {\bibfnamefont {Konstantin~S}\ \bibnamefont {Novoselov}}, \bibinfo {author}
  {\bibfnamefont {Leonid~A}\ \bibnamefont {Ponomarenko}}, \bibinfo {author}
  {\bibfnamefont {Andre~K}\ \bibnamefont {Geim}}, \ and\ \bibinfo {author}
  {\bibfnamefont {R}~\bibnamefont {Gorbachev}},\ }\bibfield  {title} {\enquote
  {\bibinfo {title} {Cross-sectional imaging of individual layers and buried
  interfaces of graphene-based heterostructures and superlattices},}\
  }\href@noop {} {\bibfield  {journal} {\bibinfo  {journal} {Nature materials}\
  }\textbf {\bibinfo {volume} {11}},\ \bibinfo {pages} {764--767} (\bibinfo
  {year} {2012})}\BibitemShut {NoStop}%
\bibitem [{\citenamefont {Rooney}\ \emph {et~al.}(2017)\citenamefont {Rooney},
  \citenamefont {Kozikov}, \citenamefont {Rudenko}, \citenamefont {Prestat},
  \citenamefont {Hamer}, \citenamefont {Withers}, \citenamefont {Cao},
  \citenamefont {Novoselov}, \citenamefont {Katsnelson}, \citenamefont
  {Gorbachev} \emph {et~al.}}]{rooney2017observing}%
  \BibitemOpen
  \bibfield  {author} {\bibinfo {author} {\bibfnamefont {Aidan~P}\ \bibnamefont
  {Rooney}}, \bibinfo {author} {\bibfnamefont {Aleksey}\ \bibnamefont
  {Kozikov}}, \bibinfo {author} {\bibfnamefont {Alexander~N}\ \bibnamefont
  {Rudenko}}, \bibinfo {author} {\bibfnamefont {Eric}\ \bibnamefont {Prestat}},
  \bibinfo {author} {\bibfnamefont {Matthew~J}\ \bibnamefont {Hamer}}, \bibinfo
  {author} {\bibfnamefont {Freddie}\ \bibnamefont {Withers}}, \bibinfo {author}
  {\bibfnamefont {Yang}\ \bibnamefont {Cao}}, \bibinfo {author} {\bibfnamefont
  {Kostya~S}\ \bibnamefont {Novoselov}}, \bibinfo {author} {\bibfnamefont
  {Mikhail~I}\ \bibnamefont {Katsnelson}}, \bibinfo {author} {\bibfnamefont
  {Roman}\ \bibnamefont {Gorbachev}},  \emph {et~al.},\ }\bibfield  {title}
  {\enquote {\bibinfo {title} {Observing imperfection in atomic interfaces for
  van der waals heterostructures},}\ }\href@noop {} {\bibfield  {journal}
  {\bibinfo  {journal} {Nano letters}\ }\textbf {\bibinfo {volume} {17}},\
  \bibinfo {pages} {5222--5228} (\bibinfo {year} {2017})}\BibitemShut {NoStop}%
\bibitem [{\citenamefont {Koma}\ \emph {et~al.}(1985)\citenamefont {Koma},
  \citenamefont {Sunouchi},\ and\ \citenamefont
  {Miyajima}}]{koma1985fabrication}%
  \BibitemOpen
  \bibfield  {author} {\bibinfo {author} {\bibfnamefont {Atsushi}\ \bibnamefont
  {Koma}}, \bibinfo {author} {\bibfnamefont {Kazumasa}\ \bibnamefont
  {Sunouchi}}, \ and\ \bibinfo {author} {\bibfnamefont {Takao}\ \bibnamefont
  {Miyajima}},\ }\bibfield  {title} {\enquote {\bibinfo {title} {Fabrication of
  ultrathin heterostructures with van der waals epitaxy},}\ }\href@noop {}
  {\bibfield  {journal} {\bibinfo  {journal} {J. Vac. Sci. Technol.}\ }\textbf
  {\bibinfo {volume} {3}},\ \bibinfo {pages} {724} (\bibinfo {year}
  {1985})}\BibitemShut {NoStop}%
\bibitem [{\citenamefont {Feldberg}\ \emph {et~al.}(2019)\citenamefont
  {Feldberg}, \citenamefont {Klymov}, \citenamefont {Garro}, \citenamefont
  {Cros}, \citenamefont {Mollard}, \citenamefont {Okuno}, \citenamefont
  {Gruart},\ and\ \citenamefont {Daudin}}]{feldberg2019spontaneous}%
  \BibitemOpen
  \bibfield  {author} {\bibinfo {author} {\bibfnamefont {Nathaniel}\
  \bibnamefont {Feldberg}}, \bibinfo {author} {\bibfnamefont {Oleksii}\
  \bibnamefont {Klymov}}, \bibinfo {author} {\bibfnamefont {Nuria}\
  \bibnamefont {Garro}}, \bibinfo {author} {\bibfnamefont {Ana}\ \bibnamefont
  {Cros}}, \bibinfo {author} {\bibfnamefont {Nicolas}\ \bibnamefont {Mollard}},
  \bibinfo {author} {\bibfnamefont {Hanako}\ \bibnamefont {Okuno}}, \bibinfo
  {author} {\bibfnamefont {Marion}\ \bibnamefont {Gruart}}, \ and\ \bibinfo
  {author} {\bibfnamefont {Bruno}\ \bibnamefont {Daudin}},\ }\bibfield  {title}
  {\enquote {\bibinfo {title} {Spontaneous intercalation of ga and in bilayers
  during plasma-assisted molecular beam epitaxy growth of gan on graphene on
  sic},}\ }\href@noop {} {\bibfield  {journal} {\bibinfo  {journal}
  {Nanotechnology}\ }\textbf {\bibinfo {volume} {30}},\ \bibinfo {pages}
  {375602} (\bibinfo {year} {2019})}\BibitemShut {NoStop}%
\bibitem [{\citenamefont {Heilmann}\ \emph {et~al.}(2018)\citenamefont
  {Heilmann}, \citenamefont {Bashouti}, \citenamefont {Riechert},\ and\
  \citenamefont {Lopes}}]{heilmann2DMat2018defect}%
  \BibitemOpen
  \bibfield  {author} {\bibinfo {author} {\bibfnamefont {M}~\bibnamefont
  {Heilmann}}, \bibinfo {author} {\bibfnamefont {M}~\bibnamefont {Bashouti}},
  \bibinfo {author} {\bibfnamefont {H}~\bibnamefont {Riechert}}, \ and\
  \bibinfo {author} {\bibfnamefont {JMJ}\ \bibnamefont {Lopes}},\ }\bibfield
  {title} {\enquote {\bibinfo {title} {Defect mediated van der waals epitaxy of
  hexagonal boron nitride on graphene},}\ }\href@noop {} {\bibfield  {journal}
  {\bibinfo  {journal} {2D Materials}\ }\textbf {\bibinfo {volume} {5}},\
  \bibinfo {pages} {025004} (\bibinfo {year} {2018})}\BibitemShut {NoStop}%
\bibitem [{\citenamefont {Heilmann}\ \emph {et~al.}(2021)\citenamefont
  {Heilmann}, \citenamefont {Deinhart}, \citenamefont {Tahraoui}, \citenamefont
  {H{\"o}flich},\ and\ \citenamefont {Lopes}}]{heilmann2021spatially}%
  \BibitemOpen
  \bibfield  {author} {\bibinfo {author} {\bibfnamefont {Martin}\ \bibnamefont
  {Heilmann}}, \bibinfo {author} {\bibfnamefont {Victor}\ \bibnamefont
  {Deinhart}}, \bibinfo {author} {\bibfnamefont {Abbes}\ \bibnamefont
  {Tahraoui}}, \bibinfo {author} {\bibfnamefont {Katja}\ \bibnamefont
  {H{\"o}flich}}, \ and\ \bibinfo {author} {\bibfnamefont {J~Marcelo~J}\
  \bibnamefont {Lopes}},\ }\bibfield  {title} {\enquote {\bibinfo {title}
  {Spatially controlled epitaxial growth of 2d heterostructures via defect
  engineering using a focused he ion beam},}\ }\href@noop {} {\bibfield
  {journal} {\bibinfo  {journal} {npj 2D Materials and Applications}\ }\textbf
  {\bibinfo {volume} {5}},\ \bibinfo {pages} {1--7} (\bibinfo {year}
  {2021})}\BibitemShut {NoStop}%
\bibitem [{\citenamefont {Wang}\ \emph {et~al.}(2021)\citenamefont {Wang},
  \citenamefont {Crowther}, \citenamefont {Kageshima}, \citenamefont {Hibino},\
  and\ \citenamefont {Taniyasu}}]{wang2021epitaxial}%
  \BibitemOpen
  \bibfield  {author} {\bibinfo {author} {\bibfnamefont {Shengnan}\
  \bibnamefont {Wang}}, \bibinfo {author} {\bibfnamefont {Jack}\ \bibnamefont
  {Crowther}}, \bibinfo {author} {\bibfnamefont {Hiroyuki}\ \bibnamefont
  {Kageshima}}, \bibinfo {author} {\bibfnamefont {Hiroki}\ \bibnamefont
  {Hibino}}, \ and\ \bibinfo {author} {\bibfnamefont {Yoshitaka}\ \bibnamefont
  {Taniyasu}},\ }\bibfield  {title} {\enquote {\bibinfo {title} {Epitaxial
  intercalation growth of scalable hexagonal boron nitride/graphene bilayer
  moir{\'e} materials with highly convergent interlayer angles},}\ }\href@noop
  {} {\bibfield  {journal} {\bibinfo  {journal} {Acs Nano}\ }\textbf {\bibinfo
  {volume} {15}},\ \bibinfo {pages} {14384--14393} (\bibinfo {year}
  {2021})}\BibitemShut {NoStop}%
\bibitem [{\citenamefont {Yang}\ \emph {et~al.}(2015)\citenamefont {Yang},
  \citenamefont {Fu}, \citenamefont {Li}, \citenamefont {Wei}, \citenamefont
  {Xiao}, \citenamefont {Wei},\ and\ \citenamefont {Bao}}]{yang2015creating}%
  \BibitemOpen
  \bibfield  {author} {\bibinfo {author} {\bibfnamefont {Yang}\ \bibnamefont
  {Yang}}, \bibinfo {author} {\bibfnamefont {Qiang}\ \bibnamefont {Fu}},
  \bibinfo {author} {\bibfnamefont {Haobo}\ \bibnamefont {Li}}, \bibinfo
  {author} {\bibfnamefont {Mingming}\ \bibnamefont {Wei}}, \bibinfo {author}
  {\bibfnamefont {Jianping}\ \bibnamefont {Xiao}}, \bibinfo {author}
  {\bibfnamefont {Wei}\ \bibnamefont {Wei}}, \ and\ \bibinfo {author}
  {\bibfnamefont {Xinhe}\ \bibnamefont {Bao}},\ }\bibfield  {title} {\enquote
  {\bibinfo {title} {Creating a nanospace under an h-bn cover for adlayer
  growth on nickel (111)},}\ }\href@noop {} {\bibfield  {journal} {\bibinfo
  {journal} {ACS nano}\ }\textbf {\bibinfo {volume} {9}},\ \bibinfo {pages}
  {11589} (\bibinfo {year} {2015})}\BibitemShut {NoStop}%
\bibitem [{\citenamefont {Al~Balushi}\ \emph {et~al.}(2016)\citenamefont
  {Al~Balushi}, \citenamefont {Wang}, \citenamefont {Ghosh}, \citenamefont
  {Vil{\'a}}, \citenamefont {Eichfeld}, \citenamefont {Caldwell}, \citenamefont
  {Qin}, \citenamefont {Lin}, \citenamefont {DeSario}, \citenamefont {Stone}
  \emph {et~al.}}]{al2016two}%
  \BibitemOpen
  \bibfield  {author} {\bibinfo {author} {\bibfnamefont {Zakaria~Y}\
  \bibnamefont {Al~Balushi}}, \bibinfo {author} {\bibfnamefont
  {Ke}~\bibnamefont {Wang}}, \bibinfo {author} {\bibfnamefont {Ram~Krishna}\
  \bibnamefont {Ghosh}}, \bibinfo {author} {\bibfnamefont {Rafael~A}\
  \bibnamefont {Vil{\'a}}}, \bibinfo {author} {\bibfnamefont {Sarah~M}\
  \bibnamefont {Eichfeld}}, \bibinfo {author} {\bibfnamefont {Joshua~D}\
  \bibnamefont {Caldwell}}, \bibinfo {author} {\bibfnamefont {Xiaoye}\
  \bibnamefont {Qin}}, \bibinfo {author} {\bibfnamefont {Yu-Chuan}\
  \bibnamefont {Lin}}, \bibinfo {author} {\bibfnamefont {Paul~A}\ \bibnamefont
  {DeSario}}, \bibinfo {author} {\bibfnamefont {Greg}\ \bibnamefont {Stone}},
  \emph {et~al.},\ }\bibfield  {title} {\enquote {\bibinfo {title}
  {Two-dimensional gallium nitride realized via graphene encapsulation},}\
  }\href@noop {} {\bibfield  {journal} {\bibinfo  {journal} {Nature materials}\
  }\textbf {\bibinfo {volume} {15}},\ \bibinfo {pages} {1166--1171} (\bibinfo
  {year} {2016})}\BibitemShut {NoStop}%
\bibitem [{\citenamefont {Lee}\ \emph {et~al.}(2022)\citenamefont {Lee},
  \citenamefont {Wang}, \citenamefont {Qin}, \citenamefont {Kim}, \citenamefont
  {Liu}, \citenamefont {Nunley}, \citenamefont {Fang}, \citenamefont
  {Maniyara}, \citenamefont {Dong}, \citenamefont {Robinson} \emph
  {et~al.}}]{lee2022confined}%
  \BibitemOpen
  \bibfield  {author} {\bibinfo {author} {\bibfnamefont {Woojoo}\ \bibnamefont
  {Lee}}, \bibinfo {author} {\bibfnamefont {Yuanxi}\ \bibnamefont {Wang}},
  \bibinfo {author} {\bibfnamefont {Wei}\ \bibnamefont {Qin}}, \bibinfo
  {author} {\bibfnamefont {Hyunsue}\ \bibnamefont {Kim}}, \bibinfo {author}
  {\bibfnamefont {Mengke}\ \bibnamefont {Liu}}, \bibinfo {author}
  {\bibfnamefont {T~Nathan}\ \bibnamefont {Nunley}}, \bibinfo {author}
  {\bibfnamefont {Bin}\ \bibnamefont {Fang}}, \bibinfo {author} {\bibfnamefont
  {Rinu}\ \bibnamefont {Maniyara}}, \bibinfo {author} {\bibfnamefont {Chengye}\
  \bibnamefont {Dong}}, \bibinfo {author} {\bibfnamefont {Joshua~A}\
  \bibnamefont {Robinson}},  \emph {et~al.},\ }\bibfield  {title} {\enquote
  {\bibinfo {title} {Confined monolayer ag as a large gap 2d semiconductor and
  its momentum resolved excited states},}\ }\href@noop {} {\bibfield  {journal}
  {\bibinfo  {journal} {Nano letters}\ }\textbf {\bibinfo {volume} {22}},\
  \bibinfo {pages} {7841--7847} (\bibinfo {year} {2022})}\BibitemShut {NoStop}%
\bibitem [{\citenamefont {Turker}\ \emph {et~al.}(2023)\citenamefont {Turker},
  \citenamefont {Dong}, \citenamefont {Wetherington}, \citenamefont
  {El-Sherif}, \citenamefont {Holoviak}, \citenamefont {Trdinich},
  \citenamefont {Lawson}, \citenamefont {Krishnan}, \citenamefont {Whittier},
  \citenamefont {Sinnott} \emph {et~al.}}]{turker20232d}%
  \BibitemOpen
  \bibfield  {author} {\bibinfo {author} {\bibfnamefont {Furkan}\ \bibnamefont
  {Turker}}, \bibinfo {author} {\bibfnamefont {Chengye}\ \bibnamefont {Dong}},
  \bibinfo {author} {\bibfnamefont {Maxwell~T}\ \bibnamefont {Wetherington}},
  \bibinfo {author} {\bibfnamefont {Hesham}\ \bibnamefont {El-Sherif}},
  \bibinfo {author} {\bibfnamefont {Stephen}\ \bibnamefont {Holoviak}},
  \bibinfo {author} {\bibfnamefont {Zachary~J}\ \bibnamefont {Trdinich}},
  \bibinfo {author} {\bibfnamefont {Eric~T}\ \bibnamefont {Lawson}}, \bibinfo
  {author} {\bibfnamefont {Gopi}\ \bibnamefont {Krishnan}}, \bibinfo {author}
  {\bibfnamefont {Caleb}\ \bibnamefont {Whittier}}, \bibinfo {author}
  {\bibfnamefont {Susan~B}\ \bibnamefont {Sinnott}},  \emph {et~al.},\
  }\bibfield  {title} {\enquote {\bibinfo {title} {2d oxides realized via
  confinement heteroepitaxy},}\ }\href@noop {} {\bibfield  {journal} {\bibinfo
  {journal} {Advanced Functional Materials}\ }\textbf {\bibinfo {volume}
  {33}},\ \bibinfo {pages} {2210404} (\bibinfo {year} {2023})}\BibitemShut
  {NoStop}%
\bibitem [{\citenamefont {Perdew}\ \emph {et~al.}(1996)\citenamefont {Perdew},
  \citenamefont {Burke},\ and\ \citenamefont {Ernzerhof}}]{PBE}%
  \BibitemOpen
  \bibfield  {author} {\bibinfo {author} {\bibfnamefont {J.~P.}\ \bibnamefont
  {Perdew}}, \bibinfo {author} {\bibfnamefont {K.}~\bibnamefont {Burke}}, \
  and\ \bibinfo {author} {\bibfnamefont {M.}~\bibnamefont {Ernzerhof}},\
  }\href@noop {} {\bibfield  {journal} {\bibinfo  {journal} {Phys. Rev. Lett.}\
  }\textbf {\bibinfo {volume} {77}},\ \bibinfo {pages} {3865} (\bibinfo {year}
  {1996})}\BibitemShut {NoStop}%
\bibitem [{\citenamefont {Bl$\rm\ddot{u}$chl}(1994)}]{paw}%
  \BibitemOpen
  \bibfield  {author} {\bibinfo {author} {\bibfnamefont {P.~E.}\ \bibnamefont
  {Bl$\rm\ddot{u}$chl}},\ }\href@noop {} {\bibfield  {journal} {\bibinfo
  {journal} {Phys. Rev. B}\ }\textbf {\bibinfo {volume} {50}},\ \bibinfo
  {pages} {17953} (\bibinfo {year} {1994})}\BibitemShut {NoStop}%
\bibitem [{\citenamefont {Kresse}\ and\ \citenamefont
  {Furthm$\rm\ddot{u}$ller}(1996{\natexlab{a}})}]{vasp1}%
  \BibitemOpen
  \bibfield  {author} {\bibinfo {author} {\bibfnamefont {G.}~\bibnamefont
  {Kresse}}\ and\ \bibinfo {author} {\bibfnamefont {J.}~\bibnamefont
  {Furthm$\rm\ddot{u}$ller}},\ }\href@noop {} {\bibfield  {journal} {\bibinfo
  {journal} {Comput. Mater. Sci.}\ }\textbf {\bibinfo {volume} {6}},\ \bibinfo
  {pages} {15} (\bibinfo {year} {1996}{\natexlab{a}})}\BibitemShut {NoStop}%
\bibitem [{\citenamefont {Kresse}\ and\ \citenamefont
  {Furthm$\rm\ddot{u}$ller}(1996{\natexlab{b}})}]{vasp2}%
  \BibitemOpen
  \bibfield  {author} {\bibinfo {author} {\bibfnamefont {G.}~\bibnamefont
  {Kresse}}\ and\ \bibinfo {author} {\bibfnamefont {J.}~\bibnamefont
  {Furthm$\rm\ddot{u}$ller}},\ }\href@noop {} {\bibfield  {journal} {\bibinfo
  {journal} {Phys. Rev. B}\ }\textbf {\bibinfo {volume} {54}},\ \bibinfo
  {pages} {11169} (\bibinfo {year} {1996}{\natexlab{b}})}\BibitemShut {NoStop}%
\bibitem [{\citenamefont {Tckatchenko}\ and\ \citenamefont
  {Scheffler}(2009)}]{tckatchenkoPRL2009}%
  \BibitemOpen
  \bibfield  {author} {\bibinfo {author} {\bibfnamefont {A.}~\bibnamefont
  {Tckatchenko}}\ and\ \bibinfo {author} {\bibfnamefont {M.}~\bibnamefont
  {Scheffler}},\ }\href@noop {} {\bibfield  {journal} {\bibinfo  {journal}
  {Phys. Rev. Lett.}\ }\textbf {\bibinfo {volume} {102}},\ \bibinfo {pages}
  {073005} (\bibinfo {year} {2009})}\BibitemShut {NoStop}%
\bibitem [{\citenamefont {Monkhorst}\ and\ \citenamefont {Pack}(1976)}]{mp}%
  \BibitemOpen
  \bibfield  {author} {\bibinfo {author} {\bibfnamefont {H.~J.}\ \bibnamefont
  {Monkhorst}}\ and\ \bibinfo {author} {\bibfnamefont {J.~D.}\ \bibnamefont
  {Pack}},\ }\href@noop {} {\bibfield  {journal} {\bibinfo  {journal} {Phys.
  Rev. B}\ }\textbf {\bibinfo {volume} {13}},\ \bibinfo {pages} {5188}
  (\bibinfo {year} {1976})}\BibitemShut {NoStop}%
\bibitem [{\citenamefont {Li}\ \emph {et~al.}(2010)\citenamefont {Li},
  \citenamefont {Luican}, \citenamefont {Lopes~dos Santos}, \citenamefont
  {Castro~Neto}, \citenamefont {Reina}, \citenamefont {Kong},\ and\
  \citenamefont {Andrei}}]{Li2010}%
  \BibitemOpen
  \bibfield  {author} {\bibinfo {author} {\bibfnamefont {Guohong}\ \bibnamefont
  {Li}}, \bibinfo {author} {\bibfnamefont {A.}~\bibnamefont {Luican}}, \bibinfo
  {author} {\bibfnamefont {J.~M.~B.}\ \bibnamefont {Lopes~dos Santos}},
  \bibinfo {author} {\bibfnamefont {A.~H.}\ \bibnamefont {Castro~Neto}},
  \bibinfo {author} {\bibfnamefont {A.}~\bibnamefont {Reina}}, \bibinfo
  {author} {\bibfnamefont {J.}~\bibnamefont {Kong}}, \ and\ \bibinfo {author}
  {\bibfnamefont {E.~Y.}\ \bibnamefont {Andrei}},\ }\bibfield  {title}
  {\enquote {\bibinfo {title} {Observation of van hove singularities in twisted
  graphene layers},}\ }\href@noop {} {\bibfield  {journal} {\bibinfo  {journal}
  {Nature Physics}\ }\textbf {\bibinfo {volume} {6}},\ \bibinfo {pages}
  {109--113} (\bibinfo {year} {2010})}\BibitemShut {NoStop}%
\bibitem [{\citenamefont {Pankratov}\ \emph {et~al.}(2010)\citenamefont
  {Pankratov}, \citenamefont {Hensel},\ and\ \citenamefont
  {Bockstedte}}]{PhysRevB.82.121416}%
  \BibitemOpen
  \bibfield  {author} {\bibinfo {author} {\bibfnamefont {O.}~\bibnamefont
  {Pankratov}}, \bibinfo {author} {\bibfnamefont {S.}~\bibnamefont {Hensel}}, \
  and\ \bibinfo {author} {\bibfnamefont {M.}~\bibnamefont {Bockstedte}},\
  }\bibfield  {title} {\enquote {\bibinfo {title} {Electron spectrum of
  epitaxial graphene monolayers},}\ }\href@noop {} {\bibfield  {journal}
  {\bibinfo  {journal} {Phys. Rev. B}\ }\textbf {\bibinfo {volume} {82}},\
  \bibinfo {pages} {121416} (\bibinfo {year} {2010})}\BibitemShut {NoStop}%
\bibitem [{\citenamefont {Padilha}\ \emph {et~al.}(2019)\citenamefont
  {Padilha}, \citenamefont {Pontes}, \citenamefont {{de Lima}}, \citenamefont
  {Kagimura},\ and\ \citenamefont {Miwa}}]{PADILHA2019603}%
  \BibitemOpen
  \bibfield  {author} {\bibinfo {author} {\bibfnamefont {J.E.}\ \bibnamefont
  {Padilha}}, \bibinfo {author} {\bibfnamefont {R.B.}\ \bibnamefont {Pontes}},
  \bibinfo {author} {\bibfnamefont {F.~Crasto}\ \bibnamefont {{de Lima}}},
  \bibinfo {author} {\bibfnamefont {R.}~\bibnamefont {Kagimura}}, \ and\
  \bibinfo {author} {\bibfnamefont {R.H.}\ \bibnamefont {Miwa}},\ }\bibfield
  {title} {\enquote {\bibinfo {title} {Graphene on the oxidized sic surface and
  the impact of the metal intercalation},}\ }\href@noop {} {\bibfield
  {journal} {\bibinfo  {journal} {Carbon}\ }\textbf {\bibinfo {volume} {145}},\
  \bibinfo {pages} {603--613} (\bibinfo {year} {2019})}\BibitemShut {NoStop}%
\bibitem [{\citenamefont {Bun{\u{a}}u}\ and\ \citenamefont
  {Calandra}(2013)}]{bunuauPRB2013}%
  \BibitemOpen
  \bibfield  {author} {\bibinfo {author} {\bibfnamefont {Oana}\ \bibnamefont
  {Bun{\u{a}}u}}\ and\ \bibinfo {author} {\bibfnamefont {Matteo}\ \bibnamefont
  {Calandra}},\ }\bibfield  {title} {\enquote {\bibinfo {title} {Projector
  augmented wave calculation of x-ray absorption spectra at the l 2, 3
  edges},}\ }\href@noop {} {\bibfield  {journal} {\bibinfo  {journal} {Physical
  Review B}\ }\textbf {\bibinfo {volume} {87}},\ \bibinfo {pages} {205105}
  (\bibinfo {year} {2013})}\BibitemShut {NoStop}%
\bibitem [{\citenamefont {Gougoussis}\ \emph {et~al.}(2009)\citenamefont
  {Gougoussis}, \citenamefont {Calandra}, \citenamefont {Seitsonen},\ and\
  \citenamefont {Mauri}}]{xas2}%
  \BibitemOpen
  \bibfield  {author} {\bibinfo {author} {\bibfnamefont {Christos}\
  \bibnamefont {Gougoussis}}, \bibinfo {author} {\bibfnamefont {Matteo}\
  \bibnamefont {Calandra}}, \bibinfo {author} {\bibfnamefont {Ari~P.}\
  \bibnamefont {Seitsonen}}, \ and\ \bibinfo {author} {\bibfnamefont
  {Francesco}\ \bibnamefont {Mauri}},\ }\bibfield  {title} {\enquote {\bibinfo
  {title} {First-principles calculations of x-ray absorption in a scheme based
  on ultrasoft pseudopotentials: From $\ensuremath{\alpha}$-quartz to
  high-${T}_{c}$ compounds},}\ }\href@noop {} {\bibfield  {journal} {\bibinfo
  {journal} {Phys. Rev. B}\ }\textbf {\bibinfo {volume} {80}},\ \bibinfo
  {pages} {075102} (\bibinfo {year} {2009})}\BibitemShut {NoStop}%
\bibitem [{\citenamefont {Taillefumier}\ \emph {et~al.}(2002)\citenamefont
  {Taillefumier}, \citenamefont {Cabaret}, \citenamefont {Flank},\ and\
  \citenamefont {Mauri}}]{taillefumierPRB2002}%
  \BibitemOpen
  \bibfield  {author} {\bibinfo {author} {\bibfnamefont {Mathieu}\ \bibnamefont
  {Taillefumier}}, \bibinfo {author} {\bibfnamefont {Delphine}\ \bibnamefont
  {Cabaret}}, \bibinfo {author} {\bibfnamefont {Anne-Marie}\ \bibnamefont
  {Flank}}, \ and\ \bibinfo {author} {\bibfnamefont {Francesco}\ \bibnamefont
  {Mauri}},\ }\bibfield  {title} {\enquote {\bibinfo {title} {X-ray absorption
  near-edge structure calculations with the pseudopotentials: Application to
  the k edge in diamond and $\alpha$-quartz},}\ }\href@noop {} {\bibfield
  {journal} {\bibinfo  {journal} {Physical Review B}\ }\textbf {\bibinfo
  {volume} {66}},\ \bibinfo {pages} {195107} (\bibinfo {year}
  {2002})}\BibitemShut {NoStop}%
\bibitem [{\citenamefont {Giannozzi{\it~et al.}}(2009)}]{espresso}%
  \BibitemOpen
  \bibfield  {author} {\bibinfo {author} {\bibfnamefont {P.}~\bibnamefont
  {Giannozzi{\it~et al.}}},\ }\href@noop {} {\bibfield  {journal} {\bibinfo
  {journal} {J. Phys.: Condens. Matter}\ }\textbf {\bibinfo {volume} {21}},\
  \bibinfo {pages} {395502} (\bibinfo {year} {2009})}\BibitemShut {NoStop}%
\bibitem [{\citenamefont {Pickard}\ and\ \citenamefont {Mauri}(2001)}]{gipaw}%
  \BibitemOpen
  \bibfield  {author} {\bibinfo {author} {\bibfnamefont {Chris~J}\ \bibnamefont
  {Pickard}}\ and\ \bibinfo {author} {\bibfnamefont {Francesco}\ \bibnamefont
  {Mauri}},\ }\bibfield  {title} {\enquote {\bibinfo {title} {All-electron
  magnetic response with pseudopotentials: Nmr chemical shifts},}\ }\href@noop
  {} {\bibfield  {journal} {\bibinfo  {journal} {Physical Review B}\ }\textbf
  {\bibinfo {volume} {63}},\ \bibinfo {pages} {245101} (\bibinfo {year}
  {2001})}\BibitemShut {NoStop}%
\bibitem [{\citenamefont {Bianchettin}\ \emph {et~al.}(2006)\citenamefont
  {Bianchettin}, \citenamefont {Baraldi}, \citenamefont {de~Gironcoli},
  \citenamefont {Lizzit}, \citenamefont {Petaccia}, \citenamefont {Vesselli},
  \citenamefont {Comelli},\ and\ \citenamefont {Rosei}}]{PhysRevB.74.045430}%
  \BibitemOpen
  \bibfield  {author} {\bibinfo {author} {\bibfnamefont {Laura}\ \bibnamefont
  {Bianchettin}}, \bibinfo {author} {\bibfnamefont {Alessandro}\ \bibnamefont
  {Baraldi}}, \bibinfo {author} {\bibfnamefont {Stefano}\ \bibnamefont
  {de~Gironcoli}}, \bibinfo {author} {\bibfnamefont {Silvano}\ \bibnamefont
  {Lizzit}}, \bibinfo {author} {\bibfnamefont {Luca}\ \bibnamefont {Petaccia}},
  \bibinfo {author} {\bibfnamefont {Erik}\ \bibnamefont {Vesselli}}, \bibinfo
  {author} {\bibfnamefont {Giovanni}\ \bibnamefont {Comelli}}, \ and\ \bibinfo
  {author} {\bibfnamefont {Renzo}\ \bibnamefont {Rosei}},\ }\bibfield  {title}
  {\enquote {\bibinfo {title} {Geometric and electronic structure of the
  nrh(100) system by core-level photoelectron spectroscopy: Experiment and
  theory},}\ }\href@noop {} {\bibfield  {journal} {\bibinfo  {journal} {Phys.
  Rev. B}\ }\textbf {\bibinfo {volume} {74}},\ \bibinfo {pages} {045430}
  (\bibinfo {year} {2006})}\BibitemShut {NoStop}%
\bibitem [{\citenamefont {Plekan}\ \emph {et~al.}(2007)\citenamefont {Plekan},
  \citenamefont {Feyer}, \citenamefont {Richter}, \citenamefont {Coreno},
  \citenamefont {de~Simone}, \citenamefont {Prince},\ and\ \citenamefont
  {Carravetta}}]{cls1}%
  \BibitemOpen
  \bibfield  {author} {\bibinfo {author} {\bibfnamefont {Oksana}\ \bibnamefont
  {Plekan}}, \bibinfo {author} {\bibfnamefont {Vitaliy}\ \bibnamefont {Feyer}},
  \bibinfo {author} {\bibfnamefont {Robert}\ \bibnamefont {Richter}}, \bibinfo
  {author} {\bibfnamefont {Marcello}\ \bibnamefont {Coreno}}, \bibinfo {author}
  {\bibfnamefont {Monica}\ \bibnamefont {de~Simone}}, \bibinfo {author}
  {\bibfnamefont {Kevin~C.}\ \bibnamefont {Prince}}, \ and\ \bibinfo {author}
  {\bibfnamefont {Vincenzo}\ \bibnamefont {Carravetta}},\ }\bibfield  {title}
  {\enquote {\bibinfo {title} {Investigation of the amino acids glycine,
  proline, and methionine by photoemission spectroscopy},}\ }\href@noop {}
  {\bibfield  {journal} {\bibinfo  {journal} {The Journal of Physical Chemistry
  A}\ }\textbf {\bibinfo {volume} {111}},\ \bibinfo {pages} {10998--11005}
  (\bibinfo {year} {2007})}\BibitemShut {NoStop}%
\bibitem [{\citenamefont {Pehlke}\ and\ \citenamefont
  {Scheffler}(1993)}]{PhysRevLett.71.2338}%
  \BibitemOpen
  \bibfield  {author} {\bibinfo {author} {\bibfnamefont {E.}~\bibnamefont
  {Pehlke}}\ and\ \bibinfo {author} {\bibfnamefont {M.}~\bibnamefont
  {Scheffler}},\ }\bibfield  {title} {\enquote {\bibinfo {title} {Evidence for
  site-sensitive screening of core holes at the si and ge (001) surface},}\
  }\href@noop {} {\bibfield  {journal} {\bibinfo  {journal} {Phys. Rev. Lett.}\
  }\textbf {\bibinfo {volume} {71}},\ \bibinfo {pages} {2338--2341} (\bibinfo
  {year} {1993})}\BibitemShut {NoStop}%
\bibitem [{\citenamefont {Bolognesi}\ \emph {et~al.}(2009)\citenamefont
  {Bolognesi}, \citenamefont {Mattioli}, \citenamefont {O’Keeffe},
  \citenamefont {Feyer}, \citenamefont {Plekan}, \citenamefont {Ovcharenko},
  \citenamefont {Prince}, \citenamefont {Coreno}, \citenamefont
  {Amore~Bonapasta},\ and\ \citenamefont
  {Avaldi}}]{bolognesi2009investigation}%
  \BibitemOpen
  \bibfield  {author} {\bibinfo {author} {\bibfnamefont {P}~\bibnamefont
  {Bolognesi}}, \bibinfo {author} {\bibfnamefont {G}~\bibnamefont {Mattioli}},
  \bibinfo {author} {\bibfnamefont {P}~\bibnamefont {O’Keeffe}}, \bibinfo
  {author} {\bibfnamefont {V}~\bibnamefont {Feyer}}, \bibinfo {author}
  {\bibfnamefont {O}~\bibnamefont {Plekan}}, \bibinfo {author} {\bibfnamefont
  {Y}~\bibnamefont {Ovcharenko}}, \bibinfo {author} {\bibfnamefont
  {KC}~\bibnamefont {Prince}}, \bibinfo {author} {\bibfnamefont
  {M}~\bibnamefont {Coreno}}, \bibinfo {author} {\bibfnamefont {A}~\bibnamefont
  {Amore~Bonapasta}}, \ and\ \bibinfo {author} {\bibfnamefont {L}~\bibnamefont
  {Avaldi}},\ }\bibfield  {title} {\enquote {\bibinfo {title} {Investigation of
  halogenated pyrimidines by x-ray photoemission spectroscopy and theoretical
  dft methods},}\ }\href@noop {} {\bibfield  {journal} {\bibinfo  {journal}
  {The Journal of Physical Chemistry A}\ }\textbf {\bibinfo {volume} {113}},\
  \bibinfo {pages} {13593--13600} (\bibinfo {year} {2009})}\BibitemShut
  {NoStop}%
\bibitem [{\citenamefont {Rondino}\ \emph {et~al.}(2014)\citenamefont
  {Rondino}, \citenamefont {Catone}, \citenamefont {Mattioli}, \citenamefont
  {Bonapasta}, \citenamefont {Bolognesi}, \citenamefont {Casavola},
  \citenamefont {Coreno}, \citenamefont {O{'}Keeffe},\ and\ \citenamefont
  {Avaldi}}]{bindingenergyBE2}%
  \BibitemOpen
  \bibfield  {author} {\bibinfo {author} {\bibfnamefont {Flaminia}\
  \bibnamefont {Rondino}}, \bibinfo {author} {\bibfnamefont {Daniele}\
  \bibnamefont {Catone}}, \bibinfo {author} {\bibfnamefont {Giuseppe}\
  \bibnamefont {Mattioli}}, \bibinfo {author} {\bibfnamefont {Aldo~Amore}\
  \bibnamefont {Bonapasta}}, \bibinfo {author} {\bibfnamefont {Paola}\
  \bibnamefont {Bolognesi}}, \bibinfo {author} {\bibfnamefont {Anna~Rita}\
  \bibnamefont {Casavola}}, \bibinfo {author} {\bibfnamefont {Marcello}\
  \bibnamefont {Coreno}}, \bibinfo {author} {\bibfnamefont {Patrick}\
  \bibnamefont {O{'}Keeffe}}, \ and\ \bibinfo {author} {\bibfnamefont
  {Lorenzo}\ \bibnamefont {Avaldi}},\ }\bibfield  {title} {\enquote {\bibinfo
  {title} {Competition between electron-donor and electron-acceptor
  substituents in nitrotoluene isomers: a photoelectron spectroscopy and ab
  initio investigation},}\ }\href@noop {} {\bibfield  {journal} {\bibinfo
  {journal} {RSC Adv.}\ }\textbf {\bibinfo {volume} {4}},\ \bibinfo {pages}
  {5272--5282} (\bibinfo {year} {2014})}\BibitemShut {NoStop}%
\bibitem [{\citenamefont {Castrovilli}\ \emph {et~al.}(2018)\citenamefont
  {Castrovilli}, \citenamefont {Bolognesi}, \citenamefont {Bodo}, \citenamefont
  {Mattioli}, \citenamefont {Cartoni},\ and\ \citenamefont
  {Avaldi}}]{castrovilli2018experimental}%
  \BibitemOpen
  \bibfield  {author} {\bibinfo {author} {\bibfnamefont {MC}~\bibnamefont
  {Castrovilli}}, \bibinfo {author} {\bibfnamefont {P}~\bibnamefont
  {Bolognesi}}, \bibinfo {author} {\bibfnamefont {E}~\bibnamefont {Bodo}},
  \bibinfo {author} {\bibfnamefont {G}~\bibnamefont {Mattioli}}, \bibinfo
  {author} {\bibfnamefont {A}~\bibnamefont {Cartoni}}, \ and\ \bibinfo {author}
  {\bibfnamefont {L}~\bibnamefont {Avaldi}},\ }\bibfield  {title} {\enquote
  {\bibinfo {title} {An experimental and theoretical investigation of xps and
  nexafs of 5-halouracils},}\ }\href@noop {} {\bibfield  {journal} {\bibinfo
  {journal} {Physical Chemistry Chemical Physics}\ }\textbf {\bibinfo {volume}
  {20}},\ \bibinfo {pages} {6657--6667} (\bibinfo {year} {2018})}\BibitemShut
  {NoStop}%
\bibitem [{\citenamefont {Larciprete}\ \emph {et~al.}(2001)\citenamefont
  {Larciprete}, \citenamefont {Goldoni}, \citenamefont {Gro{\^s}o},
  \citenamefont {Lizzit},\ and\ \citenamefont
  {Paolucci}}]{larciprete2001photochemistry}%
  \BibitemOpen
  \bibfield  {author} {\bibinfo {author} {\bibfnamefont {R}~\bibnamefont
  {Larciprete}}, \bibinfo {author} {\bibfnamefont {A}~\bibnamefont {Goldoni}},
  \bibinfo {author} {\bibfnamefont {A}~\bibnamefont {Gro{\^s}o}}, \bibinfo
  {author} {\bibfnamefont {S}~\bibnamefont {Lizzit}}, \ and\ \bibinfo {author}
  {\bibfnamefont {G}~\bibnamefont {Paolucci}},\ }\bibfield  {title} {\enquote
  {\bibinfo {title} {The photochemistry of ch4 adsorbed on pt (1 1 1) studied
  by high resolution fast xps},}\ }\href@noop {} {\bibfield  {journal}
  {\bibinfo  {journal} {Surface science}\ }\textbf {\bibinfo {volume} {482}},\
  \bibinfo {pages} {134--140} (\bibinfo {year} {2001})}\BibitemShut {NoStop}%
\bibitem [{\citenamefont {Coletti}\ \emph {et~al.}(2011)\citenamefont
  {Coletti}, \citenamefont {Emtsev}, \citenamefont {Zakharov}, \citenamefont
  {Ouisse}, \citenamefont {Chaussende},\ and\ \citenamefont
  {Starke}}]{coletti2011large}%
  \BibitemOpen
  \bibfield  {author} {\bibinfo {author} {\bibfnamefont {Camilla}\ \bibnamefont
  {Coletti}}, \bibinfo {author} {\bibfnamefont {Konstantin~V}\ \bibnamefont
  {Emtsev}}, \bibinfo {author} {\bibfnamefont {Alexei~A}\ \bibnamefont
  {Zakharov}}, \bibinfo {author} {\bibfnamefont {Thierry}\ \bibnamefont
  {Ouisse}}, \bibinfo {author} {\bibfnamefont {Didier}\ \bibnamefont
  {Chaussende}}, \ and\ \bibinfo {author} {\bibfnamefont {Ulrich}\ \bibnamefont
  {Starke}},\ }\bibfield  {title} {\enquote {\bibinfo {title} {Large area
  quasi-free standing monolayer graphene on 3c-sic (111)},}\ }\href@noop {}
  {\bibfield  {journal} {\bibinfo  {journal} {Applied Physics Letters}\
  }\textbf {\bibinfo {volume} {99}},\ \bibinfo {pages} {081904} (\bibinfo
  {year} {2011})}\BibitemShut {NoStop}%
\bibitem [{\citenamefont {Giovannetti}\ \emph {et~al.}(2007)\citenamefont
  {Giovannetti}, \citenamefont {Khomyakov}, \citenamefont {Brocks},
  \citenamefont {Kelly},\ and\ \citenamefont {Van
  Den~Brink}}]{giovannettiPRB2007substrate}%
  \BibitemOpen
  \bibfield  {author} {\bibinfo {author} {\bibfnamefont {Gianluca}\
  \bibnamefont {Giovannetti}}, \bibinfo {author} {\bibfnamefont {Petr~A}\
  \bibnamefont {Khomyakov}}, \bibinfo {author} {\bibfnamefont {Geert}\
  \bibnamefont {Brocks}}, \bibinfo {author} {\bibfnamefont {Paul~J}\
  \bibnamefont {Kelly}}, \ and\ \bibinfo {author} {\bibfnamefont {Jeroen}\
  \bibnamefont {Van Den~Brink}},\ }\bibfield  {title} {\enquote {\bibinfo
  {title} {Substrate-induced band gap in graphene on hexagonal boron nitride:
  Ab initio density functional calculations},}\ }\href@noop {} {\bibfield
  {journal} {\bibinfo  {journal} {Physical Review B}\ }\textbf {\bibinfo
  {volume} {76}},\ \bibinfo {pages} {073103} (\bibinfo {year}
  {2007})}\BibitemShut {NoStop}%
\bibitem [{\citenamefont {Fan}\ \emph {et~al.}(2011)\citenamefont {Fan},
  \citenamefont {Zhao}, \citenamefont {Wang}, \citenamefont {Zhang},\ and\
  \citenamefont {Zhang}}]{fan2011tunable}%
  \BibitemOpen
  \bibfield  {author} {\bibinfo {author} {\bibfnamefont {Yingcai}\ \bibnamefont
  {Fan}}, \bibinfo {author} {\bibfnamefont {Mingwen}\ \bibnamefont {Zhao}},
  \bibinfo {author} {\bibfnamefont {Zhenhai}\ \bibnamefont {Wang}}, \bibinfo
  {author} {\bibfnamefont {Xuejuan}\ \bibnamefont {Zhang}}, \ and\ \bibinfo
  {author} {\bibfnamefont {Hongyu}\ \bibnamefont {Zhang}},\ }\bibfield  {title}
  {\enquote {\bibinfo {title} {Tunable electronic structures of graphene/boron
  nitride heterobilayers},}\ }\href@noop {} {\bibfield  {journal} {\bibinfo
  {journal} {Applied Physics Letters}\ }\textbf {\bibinfo {volume} {98}},\
  \bibinfo {pages} {083103} (\bibinfo {year} {2011})}\BibitemShut {NoStop}%
\bibitem [{\citenamefont {Kim}\ \emph {et~al.}(2013)\citenamefont {Kim},
  \citenamefont {Hsu}, \citenamefont {Araujo}, \citenamefont {Lee},
  \citenamefont {Palacios}, \citenamefont {Dresselhaus}, \citenamefont
  {Idrobo}, \citenamefont {Kim},\ and\ \citenamefont
  {Kong}}]{kim2013synthesis}%
  \BibitemOpen
  \bibfield  {author} {\bibinfo {author} {\bibfnamefont {Soo~Min}\ \bibnamefont
  {Kim}}, \bibinfo {author} {\bibfnamefont {Allen}\ \bibnamefont {Hsu}},
  \bibinfo {author} {\bibfnamefont {PT}~\bibnamefont {Araujo}}, \bibinfo
  {author} {\bibfnamefont {Yi-Hsien}\ \bibnamefont {Lee}}, \bibinfo {author}
  {\bibfnamefont {Tomas}\ \bibnamefont {Palacios}}, \bibinfo {author}
  {\bibfnamefont {Mildred}\ \bibnamefont {Dresselhaus}}, \bibinfo {author}
  {\bibfnamefont {Juan-Carlos}\ \bibnamefont {Idrobo}}, \bibinfo {author}
  {\bibfnamefont {Ki~Kang}\ \bibnamefont {Kim}}, \ and\ \bibinfo {author}
  {\bibfnamefont {Jing}\ \bibnamefont {Kong}},\ }\bibfield  {title} {\enquote
  {\bibinfo {title} {Synthesis of patched or stacked graphene and hbn flakes: a
  route to hybrid structure discovery},}\ }\href@noop {} {\bibfield  {journal}
  {\bibinfo  {journal} {Nano letters}\ }\textbf {\bibinfo {volume} {13}},\
  \bibinfo {pages} {933--941} (\bibinfo {year} {2013})}\BibitemShut {NoStop}%
\bibitem [{\citenamefont {Pan}\ \emph {et~al.}(2016)\citenamefont {Pan},
  \citenamefont {Liang}, \citenamefont {Lin}, \citenamefont {Kim},
  \citenamefont {Li}, \citenamefont {Kong}, \citenamefont {Dresselhaus},\ and\
  \citenamefont {Meunier}}]{pan2016modification}%
  \BibitemOpen
  \bibfield  {author} {\bibinfo {author} {\bibfnamefont {Minghu}\ \bibnamefont
  {Pan}}, \bibinfo {author} {\bibfnamefont {Liangbo}\ \bibnamefont {Liang}},
  \bibinfo {author} {\bibfnamefont {Wenzhi}\ \bibnamefont {Lin}}, \bibinfo
  {author} {\bibfnamefont {Soo~Min}\ \bibnamefont {Kim}}, \bibinfo {author}
  {\bibfnamefont {Qing}\ \bibnamefont {Li}}, \bibinfo {author} {\bibfnamefont
  {Jing}\ \bibnamefont {Kong}}, \bibinfo {author} {\bibfnamefont {Mildred~S}\
  \bibnamefont {Dresselhaus}}, \ and\ \bibinfo {author} {\bibfnamefont
  {Vincent}\ \bibnamefont {Meunier}},\ }\bibfield  {title} {\enquote {\bibinfo
  {title} {Modification of the electronic properties of hexagonal boron-nitride
  in bn/graphene vertical heterostructures},}\ }\href@noop {} {\bibfield
  {journal} {\bibinfo  {journal} {2D Materials}\ }\textbf {\bibinfo {volume}
  {3}},\ \bibinfo {pages} {045002} (\bibinfo {year} {2016})}\BibitemShut
  {NoStop}%
\bibitem [{\citenamefont {Wang}\ \emph {et~al.}(2017)\citenamefont {Wang},
  \citenamefont {Ma},\ and\ \citenamefont {Sun}}]{wang2017graphene}%
  \BibitemOpen
  \bibfield  {author} {\bibinfo {author} {\bibfnamefont {Jingang}\ \bibnamefont
  {Wang}}, \bibinfo {author} {\bibfnamefont {Fengcai}\ \bibnamefont {Ma}}, \
  and\ \bibinfo {author} {\bibfnamefont {Mengtao}\ \bibnamefont {Sun}},\
  }\bibfield  {title} {\enquote {\bibinfo {title} {Graphene, hexagonal boron
  nitride, and their heterostructures: properties and applications},}\
  }\href@noop {} {\bibfield  {journal} {\bibinfo  {journal} {RSC advances}\
  }\textbf {\bibinfo {volume} {7}},\ \bibinfo {pages} {16801--16822} (\bibinfo
  {year} {2017})}\BibitemShut {NoStop}%
\bibitem [{\citenamefont {Indolese}\ \emph {et~al.}(2018)\citenamefont
  {Indolese}, \citenamefont {Delagrange}, \citenamefont {Makk}, \citenamefont
  {Wallbank}, \citenamefont {Wanatabe}, \citenamefont {Taniguchi},\ and\
  \citenamefont {Sch{\"o}nenberger}}]{indolese2018signatures}%
  \BibitemOpen
  \bibfield  {author} {\bibinfo {author} {\bibfnamefont {DI}~\bibnamefont
  {Indolese}}, \bibinfo {author} {\bibfnamefont {R}~\bibnamefont {Delagrange}},
  \bibinfo {author} {\bibfnamefont {P}~\bibnamefont {Makk}}, \bibinfo {author}
  {\bibfnamefont {JR}~\bibnamefont {Wallbank}}, \bibinfo {author}
  {\bibfnamefont {K}~\bibnamefont {Wanatabe}}, \bibinfo {author} {\bibfnamefont
  {T}~\bibnamefont {Taniguchi}}, \ and\ \bibinfo {author} {\bibfnamefont
  {C}~\bibnamefont {Sch{\"o}nenberger}},\ }\bibfield  {title} {\enquote
  {\bibinfo {title} {Signatures of van hove singularities probed by the
  supercurrent in a graphene-hbn superlattice},}\ }\href@noop {} {\bibfield
  {journal} {\bibinfo  {journal} {Physical review letters}\ }\textbf {\bibinfo
  {volume} {121}},\ \bibinfo {pages} {137701} (\bibinfo {year}
  {2018})}\BibitemShut {NoStop}%
\bibitem [{\citenamefont {Torres-Rojas}\ \emph {et~al.}(2022)\citenamefont
  {Torres-Rojas}, \citenamefont {Contreras-Solorio}, \citenamefont
  {Hern{\'a}ndez},\ and\ \citenamefont {Enciso}}]{torres2022band}%
  \BibitemOpen
  \bibfield  {author} {\bibinfo {author} {\bibfnamefont {Ra{\'u}l~M}\
  \bibnamefont {Torres-Rojas}}, \bibinfo {author} {\bibfnamefont {David~A}\
  \bibnamefont {Contreras-Solorio}}, \bibinfo {author} {\bibfnamefont {Luis}\
  \bibnamefont {Hern{\'a}ndez}}, \ and\ \bibinfo {author} {\bibfnamefont
  {Agust{\'\i}n}\ \bibnamefont {Enciso}},\ }\bibfield  {title} {\enquote
  {\bibinfo {title} {Band gap variation in bi, tri and few-layered 2d
  graphene/hbn heterostructures},}\ }\href@noop {} {\bibfield  {journal}
  {\bibinfo  {journal} {Solid State Communications}\ }\textbf {\bibinfo
  {volume} {341}},\ \bibinfo {pages} {114553} (\bibinfo {year}
  {2022})}\BibitemShut {NoStop}%
\bibitem [{\citenamefont {Ramasubramaniam}\ \emph {et~al.}(2011)\citenamefont
  {Ramasubramaniam}, \citenamefont {Naveh},\ and\ \citenamefont
  {Towe}}]{ramasubramaniam2011tunable}%
  \BibitemOpen
  \bibfield  {author} {\bibinfo {author} {\bibfnamefont {Ashwin}\ \bibnamefont
  {Ramasubramaniam}}, \bibinfo {author} {\bibfnamefont {Doron}\ \bibnamefont
  {Naveh}}, \ and\ \bibinfo {author} {\bibfnamefont {Elias}\ \bibnamefont
  {Towe}},\ }\bibfield  {title} {\enquote {\bibinfo {title} {Tunable band gaps
  in bilayer graphene- bn heterostructures},}\ }\href@noop {} {\bibfield
  {journal} {\bibinfo  {journal} {Nano letters}\ }\textbf {\bibinfo {volume}
  {11}},\ \bibinfo {pages} {1070--1075} (\bibinfo {year} {2011})}\BibitemShut
  {NoStop}%
\bibitem [{\citenamefont {Mammadov}\ \emph {et~al.}(2017)\citenamefont
  {Mammadov}, \citenamefont {Ristein}, \citenamefont {Krone}, \citenamefont
  {Raidel}, \citenamefont {Wanke}, \citenamefont {Wiesmann}, \citenamefont
  {Speck},\ and\ \citenamefont {Seyller}}]{mammadov2017work}%
  \BibitemOpen
  \bibfield  {author} {\bibinfo {author} {\bibfnamefont {Samir}\ \bibnamefont
  {Mammadov}}, \bibinfo {author} {\bibfnamefont {J{\"u}rgen}\ \bibnamefont
  {Ristein}}, \bibinfo {author} {\bibfnamefont {Julia}\ \bibnamefont {Krone}},
  \bibinfo {author} {\bibfnamefont {Christian}\ \bibnamefont {Raidel}},
  \bibinfo {author} {\bibfnamefont {Martina}\ \bibnamefont {Wanke}}, \bibinfo
  {author} {\bibfnamefont {Veit}\ \bibnamefont {Wiesmann}}, \bibinfo {author}
  {\bibfnamefont {Florian}\ \bibnamefont {Speck}}, \ and\ \bibinfo {author}
  {\bibfnamefont {Thomas}\ \bibnamefont {Seyller}},\ }\bibfield  {title}
  {\enquote {\bibinfo {title} {Work function of graphene multilayers on sic
  (0001)},}\ }\href@noop {} {\bibfield  {journal} {\bibinfo  {journal} {2D
  Materials}\ }\textbf {\bibinfo {volume} {4}},\ \bibinfo {pages} {015043}
  (\bibinfo {year} {2017})}\BibitemShut {NoStop}%
\bibitem [{\citenamefont {Mattausch}\ and\ \citenamefont
  {Pankratov}(2007)}]{mattauschPRL2007}%
  \BibitemOpen
  \bibfield  {author} {\bibinfo {author} {\bibfnamefont {Alexander}\
  \bibnamefont {Mattausch}}\ and\ \bibinfo {author} {\bibfnamefont {Oleg}\
  \bibnamefont {Pankratov}},\ }\bibfield  {title} {\enquote {\bibinfo {title}
  {Ab initio study of graphene on sic},}\ }\href@noop {} {\bibfield  {journal}
  {\bibinfo  {journal} {Physical Review Letters}\ }\textbf {\bibinfo {volume}
  {99}},\ \bibinfo {pages} {076802} (\bibinfo {year} {2007})}\BibitemShut
  {NoStop}%
\bibitem [{\citenamefont {Varchon}\ \emph {et~al.}(2007)\citenamefont
  {Varchon}, \citenamefont {Feng}, \citenamefont {Hass}, \citenamefont {Li},
  \citenamefont {Nguyen}, \citenamefont {Naud}, \citenamefont {Mallet},
  \citenamefont {Veuillen}, \citenamefont {Berger}, \citenamefont {Conrad},\
  and\ \citenamefont {Magaud}}]{varchonPRL2007}%
  \BibitemOpen
  \bibfield  {author} {\bibinfo {author} {\bibfnamefont {F.}~\bibnamefont
  {Varchon}}, \bibinfo {author} {\bibfnamefont {R.}~\bibnamefont {Feng}},
  \bibinfo {author} {\bibfnamefont {J.}~\bibnamefont {Hass}}, \bibinfo {author}
  {\bibfnamefont {X.}~\bibnamefont {Li}}, \bibinfo {author} {\bibfnamefont
  {B.~Ngoc}\ \bibnamefont {Nguyen}}, \bibinfo {author} {\bibfnamefont
  {C.}~\bibnamefont {Naud}}, \bibinfo {author} {\bibfnamefont {P.}~\bibnamefont
  {Mallet}}, \bibinfo {author} {\bibfnamefont {J.-Y.}\ \bibnamefont
  {Veuillen}}, \bibinfo {author} {\bibfnamefont {C.}~\bibnamefont {Berger}},
  \bibinfo {author} {\bibfnamefont {E.~H.}\ \bibnamefont {Conrad}}, \ and\
  \bibinfo {author} {\bibfnamefont {L.}~\bibnamefont {Magaud}},\ }\href@noop {}
  {\bibfield  {journal} {\bibinfo  {journal} {Phys. Rev. Lett.}\ }\textbf
  {\bibinfo {volume} {99}},\ \bibinfo {pages} {126805} (\bibinfo {year}
  {2007})}\BibitemShut {NoStop}%
\bibitem [{\citenamefont {Pankratov}\ \emph {et~al.}(2012)\citenamefont
  {Pankratov}, \citenamefont {Hensel}, \citenamefont {G$\rm\ddot{o}$tzfried},\
  and\ \citenamefont {Bockstedte}}]{pankratov2012}%
  \BibitemOpen
  \bibfield  {author} {\bibinfo {author} {\bibfnamefont {O.}~\bibnamefont
  {Pankratov}}, \bibinfo {author} {\bibfnamefont {S.}~\bibnamefont {Hensel}},
  \bibinfo {author} {\bibfnamefont {P.}~\bibnamefont {G$\rm\ddot{o}$tzfried}},
  \ and\ \bibinfo {author} {\bibfnamefont {M.}~\bibnamefont {Bockstedte}},\
  }\href@noop {} {\bibfield  {journal} {\bibinfo  {journal} {Phys. Rev. B}\
  }\textbf {\bibinfo {volume} {86}},\ \bibinfo {pages} {155432} (\bibinfo
  {year} {2012})}\BibitemShut {NoStop}%
\bibitem [{\citenamefont {Sclauzero}\ and\ \citenamefont
  {Pasquarello}(2012)}]{sclauzeroPRB2012}%
  \BibitemOpen
  \bibfield  {author} {\bibinfo {author} {\bibfnamefont {Gabriele}\
  \bibnamefont {Sclauzero}}\ and\ \bibinfo {author} {\bibfnamefont {Alfredo}\
  \bibnamefont {Pasquarello}},\ }\href@noop {} {\bibfield  {journal} {\bibinfo
  {journal} {Phys. Rev. B}\ }\textbf {\bibinfo {volume} {85}},\ \bibinfo
  {pages} {161405(R)} (\bibinfo {year} {2012})}\BibitemShut {NoStop}%
\bibitem [{\citenamefont {Zhou}\ \emph {et~al.}(2007)\citenamefont {Zhou},
  \citenamefont {Gweon}, \citenamefont {Fedorov}, \citenamefont {First},
  \citenamefont {Heer}, \citenamefont {Guinea}, \citenamefont {Neto},\ and\
  \citenamefont {Lamzara}}]{zhouNatMat2007}%
  \BibitemOpen
  \bibfield  {author} {\bibinfo {author} {\bibfnamefont {S.~Y.}\ \bibnamefont
  {Zhou}}, \bibinfo {author} {\bibfnamefont {{G.-H.}}\ \bibnamefont {Gweon}},
  \bibinfo {author} {\bibfnamefont {A.~V.}\ \bibnamefont {Fedorov}}, \bibinfo
  {author} {\bibfnamefont {P.~N.}\ \bibnamefont {First}}, \bibinfo {author}
  {\bibfnamefont {W.~A.~DE}\ \bibnamefont {Heer}}, \bibinfo {author}
  {\bibfnamefont {{D.-H.} Lee~F.}\ \bibnamefont {Guinea}}, \bibinfo {author}
  {\bibfnamefont {A.~H.~Castro}\ \bibnamefont {Neto}}, \ and\ \bibinfo {author}
  {\bibfnamefont {A.}~\bibnamefont {Lamzara}},\ }\href@noop {} {\bibfield
  {journal} {\bibinfo  {journal} {Nature Material.}\ }\textbf {\bibinfo
  {volume} {6}},\ \bibinfo {pages} {770} (\bibinfo {year} {2007})}\BibitemShut
  {NoStop}%
\bibitem [{Note1()}]{Note1}%
  \BibitemOpen
  \bibinfo {note} {Here we have defined $\Delta \rho $ as $\rho _\protect \text
  {G/h-BN/SiC}$-($\rho _\protect \text {G/h-BN}$+$\rho _\protect \text {SiC}$)
  in G/h-BN/SiC, and $\rho _\protect \text {h-BN/G/SiC}$-$(\rho _\protect \text
  {h-BN/G}$+$\rho _\protect \text {SiC}$) in h-BN/G/SiC.}\BibitemShut {Stop}%
\bibitem [{\citenamefont {Yu}\ \emph {et~al.}(2009)\citenamefont {Yu},
  \citenamefont {Zhao}, \citenamefont {Ryu}, \citenamefont {Brus},
  \citenamefont {Kim},\ and\ \citenamefont {Kim}}]{yu2009tuning}%
  \BibitemOpen
  \bibfield  {author} {\bibinfo {author} {\bibfnamefont {Young-Jun}\
  \bibnamefont {Yu}}, \bibinfo {author} {\bibfnamefont {Yue}\ \bibnamefont
  {Zhao}}, \bibinfo {author} {\bibfnamefont {Sunmin}\ \bibnamefont {Ryu}},
  \bibinfo {author} {\bibfnamefont {Louis~E}\ \bibnamefont {Brus}}, \bibinfo
  {author} {\bibfnamefont {Kwang~S}\ \bibnamefont {Kim}}, \ and\ \bibinfo
  {author} {\bibfnamefont {Philip}\ \bibnamefont {Kim}},\ }\bibfield  {title}
  {\enquote {\bibinfo {title} {Tuning the graphene work function by electric
  field effect},}\ }\href@noop {} {\bibfield  {journal} {\bibinfo  {journal}
  {Nano letters}\ }\textbf {\bibinfo {volume} {9}},\ \bibinfo {pages}
  {3430--3434} (\bibinfo {year} {2009})}\BibitemShut {NoStop}%
\bibitem [{\citenamefont {Shtepliuk}\ \emph {et~al.}(2017)\citenamefont
  {Shtepliuk}, \citenamefont {Iakimov}, \citenamefont {Khranovskyy},
  \citenamefont {Eriksson}, \citenamefont {Giannazzo},\ and\ \citenamefont
  {Yakimova}}]{shtepliuk2017role}%
  \BibitemOpen
  \bibfield  {author} {\bibinfo {author} {\bibfnamefont {Ivan}\ \bibnamefont
  {Shtepliuk}}, \bibinfo {author} {\bibfnamefont {Tihomir}\ \bibnamefont
  {Iakimov}}, \bibinfo {author} {\bibfnamefont {Volodymyr}\ \bibnamefont
  {Khranovskyy}}, \bibinfo {author} {\bibfnamefont {Jens}\ \bibnamefont
  {Eriksson}}, \bibinfo {author} {\bibfnamefont {Filippo}\ \bibnamefont
  {Giannazzo}}, \ and\ \bibinfo {author} {\bibfnamefont {Rositsa}\ \bibnamefont
  {Yakimova}},\ }\bibfield  {title} {\enquote {\bibinfo {title} {Role of the
  potential barrier in the electrical performance of the graphene/sic
  interface},}\ }\href@noop {} {\bibfield  {journal} {\bibinfo  {journal}
  {Crystals}\ }\textbf {\bibinfo {volume} {7}},\ \bibinfo {pages} {162}
  (\bibinfo {year} {2017})}\BibitemShut {NoStop}%
\bibitem [{\citenamefont {Lopes}(2021)}]{lopesProgCrystGrow2021synthesis}%
  \BibitemOpen
  \bibfield  {author} {\bibinfo {author} {\bibfnamefont {J~Marcelo~J}\
  \bibnamefont {Lopes}},\ }\bibfield  {title} {\enquote {\bibinfo {title}
  {Synthesis of hexagonal boron nitride: From bulk crystals to atomically thin
  films},}\ }\href@noop {} {\bibfield  {journal} {\bibinfo  {journal} {Progress
  in Crystal Growth and Characterization of Materials}\ }\textbf {\bibinfo
  {volume} {67}},\ \bibinfo {pages} {100522} (\bibinfo {year}
  {2021})}\BibitemShut {NoStop}%
\bibitem [{\citenamefont {Schiros}\ \emph {et~al.}(2012)\citenamefont
  {Schiros}, \citenamefont {Nordlund}, \citenamefont {Palova}, \citenamefont
  {Prezzi}, \citenamefont {Zhao}, \citenamefont {Kim}, \citenamefont
  {Wurstbauer}, \citenamefont {Gutierrez}, \citenamefont {Delongchamp},
  \citenamefont {Jaye} \emph {et~al.}}]{schiros2012connecting}%
  \BibitemOpen
  \bibfield  {author} {\bibinfo {author} {\bibfnamefont {Theanne}\ \bibnamefont
  {Schiros}}, \bibinfo {author} {\bibfnamefont {Dennis}\ \bibnamefont
  {Nordlund}}, \bibinfo {author} {\bibfnamefont {Lucia}\ \bibnamefont
  {Palova}}, \bibinfo {author} {\bibfnamefont {Deborah}\ \bibnamefont
  {Prezzi}}, \bibinfo {author} {\bibfnamefont {Liuyan}\ \bibnamefont {Zhao}},
  \bibinfo {author} {\bibfnamefont {Keun~Soo}\ \bibnamefont {Kim}}, \bibinfo
  {author} {\bibfnamefont {Ulrich}\ \bibnamefont {Wurstbauer}}, \bibinfo
  {author} {\bibfnamefont {Christopher}\ \bibnamefont {Gutierrez}}, \bibinfo
  {author} {\bibfnamefont {Dean}\ \bibnamefont {Delongchamp}}, \bibinfo
  {author} {\bibfnamefont {Cherno}\ \bibnamefont {Jaye}},  \emph {et~al.},\
  }\bibfield  {title} {\enquote {\bibinfo {title} {Connecting dopant bond type
  with electronic structure in n-doped graphene},}\ }\href@noop {} {\bibfield
  {journal} {\bibinfo  {journal} {Nano letters}\ }\textbf {\bibinfo {volume}
  {12}},\ \bibinfo {pages} {4025--4031} (\bibinfo {year} {2012})}\BibitemShut
  {NoStop}%
\bibitem [{\citenamefont {Ouerghi}\ \emph {et~al.}(2012)\citenamefont
  {Ouerghi}, \citenamefont {Silly}, \citenamefont {Marangolo}, \citenamefont
  {Mathieu}, \citenamefont {Eddrief}, \citenamefont {Picher}, \citenamefont
  {Sirotti}, \citenamefont {El~Moussaoui},\ and\ \citenamefont
  {Belkhou}}]{ouerghi2012large}%
  \BibitemOpen
  \bibfield  {author} {\bibinfo {author} {\bibfnamefont {Abdelkarim}\
  \bibnamefont {Ouerghi}}, \bibinfo {author} {\bibfnamefont {Mathieu~G}\
  \bibnamefont {Silly}}, \bibinfo {author} {\bibfnamefont {Massimiliano}\
  \bibnamefont {Marangolo}}, \bibinfo {author} {\bibfnamefont {Claire}\
  \bibnamefont {Mathieu}}, \bibinfo {author} {\bibfnamefont {Mahmoud}\
  \bibnamefont {Eddrief}}, \bibinfo {author} {\bibfnamefont {Matthieu}\
  \bibnamefont {Picher}}, \bibinfo {author} {\bibfnamefont {Fausto}\
  \bibnamefont {Sirotti}}, \bibinfo {author} {\bibfnamefont {Souliman}\
  \bibnamefont {El~Moussaoui}}, \ and\ \bibinfo {author} {\bibfnamefont
  {Rachid}\ \bibnamefont {Belkhou}},\ }\bibfield  {title} {\enquote {\bibinfo
  {title} {Large-area and high-quality epitaxial graphene on off-axis sic
  wafers},}\ }\href@noop {} {\bibfield  {journal} {\bibinfo  {journal} {Acs
  Nano}\ }\textbf {\bibinfo {volume} {6}},\ \bibinfo {pages} {6075--6082}
  (\bibinfo {year} {2012})}\BibitemShut {NoStop}%
\bibitem [{\citenamefont {Lippitz}\ \emph {et~al.}(2013)\citenamefont
  {Lippitz}, \citenamefont {Friedrich},\ and\ \citenamefont
  {Unger}}]{lippitz2013plasma}%
  \BibitemOpen
  \bibfield  {author} {\bibinfo {author} {\bibfnamefont {Andreas}\ \bibnamefont
  {Lippitz}}, \bibinfo {author} {\bibfnamefont {J{\"o}rg~F}\ \bibnamefont
  {Friedrich}}, \ and\ \bibinfo {author} {\bibfnamefont {Wolfgang~ES}\
  \bibnamefont {Unger}},\ }\bibfield  {title} {\enquote {\bibinfo {title}
  {Plasma bromination of hopg surfaces: A nexafs and synchrotron xps study},}\
  }\href@noop {} {\bibfield  {journal} {\bibinfo  {journal} {Surface science}\
  }\textbf {\bibinfo {volume} {611}},\ \bibinfo {pages} {L1--L7} (\bibinfo
  {year} {2013})}\BibitemShut {NoStop}%
\bibitem [{\citenamefont {Greczynski}\ and\ \citenamefont
  {Hultman}(2020)}]{greczynski2020x}%
  \BibitemOpen
  \bibfield  {author} {\bibinfo {author} {\bibfnamefont {Grzegorz}\
  \bibnamefont {Greczynski}}\ and\ \bibinfo {author} {\bibfnamefont {Lars}\
  \bibnamefont {Hultman}},\ }\bibfield  {title} {\enquote {\bibinfo {title}
  {X-ray photoelectron spectroscopy: towards reliable binding energy
  referencing},}\ }\href@noop {} {\bibfield  {journal} {\bibinfo  {journal}
  {Progress in Materials Science}\ }\textbf {\bibinfo {volume} {107}},\
  \bibinfo {pages} {100591} (\bibinfo {year} {2020})}\BibitemShut {NoStop}%
\bibitem [{\citenamefont {Sediri}\ \emph {et~al.}(2015)\citenamefont {Sediri},
  \citenamefont {Pierucci}, \citenamefont {Hajlaoui}, \citenamefont {Henck},
  \citenamefont {Patriarche}, \citenamefont {Dappe}, \citenamefont {Yuan},
  \citenamefont {Toury}, \citenamefont {Belkhou}, \citenamefont {Silly} \emph
  {et~al.}}]{sediriSciRep2015atomically}%
  \BibitemOpen
  \bibfield  {author} {\bibinfo {author} {\bibfnamefont {Haikel}\ \bibnamefont
  {Sediri}}, \bibinfo {author} {\bibfnamefont {Debora}\ \bibnamefont
  {Pierucci}}, \bibinfo {author} {\bibfnamefont {Mahdi}\ \bibnamefont
  {Hajlaoui}}, \bibinfo {author} {\bibfnamefont {Hugo}\ \bibnamefont {Henck}},
  \bibinfo {author} {\bibfnamefont {Gilles}\ \bibnamefont {Patriarche}},
  \bibinfo {author} {\bibfnamefont {Yannick~J}\ \bibnamefont {Dappe}}, \bibinfo
  {author} {\bibfnamefont {Sheng}\ \bibnamefont {Yuan}}, \bibinfo {author}
  {\bibfnamefont {B{\'e}rang{\`e}re}\ \bibnamefont {Toury}}, \bibinfo {author}
  {\bibfnamefont {Rachid}\ \bibnamefont {Belkhou}}, \bibinfo {author}
  {\bibfnamefont {Mathieu~G}\ \bibnamefont {Silly}},  \emph {et~al.},\
  }\bibfield  {title} {\enquote {\bibinfo {title} {Atomically sharp interface
  in an h-bn-epitaxial graphene van der waals heterostructure},}\ }\href@noop
  {} {\bibfield  {journal} {\bibinfo  {journal} {Scientific reports}\ }\textbf
  {\bibinfo {volume} {5}},\ \bibinfo {pages} {1--10} (\bibinfo {year}
  {2015})}\BibitemShut {NoStop}%
\bibitem [{\citenamefont {Ugolotti}\ \emph {et~al.}(2017)\citenamefont
  {Ugolotti}, \citenamefont {Harivyasi}, \citenamefont {Baby}, \citenamefont
  {Dominguez}, \citenamefont {Pinardi}, \citenamefont {Lopez}, \citenamefont
  {Martin-Gago}, \citenamefont {Fratesi}, \citenamefont {Floreano},\ and\
  \citenamefont {Brivio}}]{ugolotti2017chemisorption}%
  \BibitemOpen
  \bibfield  {author} {\bibinfo {author} {\bibfnamefont {Aldo}\ \bibnamefont
  {Ugolotti}}, \bibinfo {author} {\bibfnamefont {Shashank~S}\ \bibnamefont
  {Harivyasi}}, \bibinfo {author} {\bibfnamefont {Anu}\ \bibnamefont {Baby}},
  \bibinfo {author} {\bibfnamefont {Marcos}\ \bibnamefont {Dominguez}},
  \bibinfo {author} {\bibfnamefont {Anna~Lisa}\ \bibnamefont {Pinardi}},
  \bibinfo {author} {\bibfnamefont {Maria~Francisca}\ \bibnamefont {Lopez}},
  \bibinfo {author} {\bibfnamefont {Jose~Angel}\ \bibnamefont {Martin-Gago}},
  \bibinfo {author} {\bibfnamefont {Guido}\ \bibnamefont {Fratesi}}, \bibinfo
  {author} {\bibfnamefont {Luca}\ \bibnamefont {Floreano}}, \ and\ \bibinfo
  {author} {\bibfnamefont {Gian~Paolo}\ \bibnamefont {Brivio}},\ }\bibfield
  {title} {\enquote {\bibinfo {title} {Chemisorption of pentacene on pt (111)
  with a little molecular distortion},}\ }\href@noop {} {\bibfield  {journal}
  {\bibinfo  {journal} {The Journal of Physical Chemistry C}\ }\textbf
  {\bibinfo {volume} {121}},\ \bibinfo {pages} {22797--22805} (\bibinfo {year}
  {2017})}\BibitemShut {NoStop}%
\bibitem [{\citenamefont {Golze}\ \emph {et~al.}(2022)\citenamefont {Golze},
  \citenamefont {Hirvensalo}, \citenamefont {Hern{\'a}ndez-Le{\'o}n},
  \citenamefont {Aarva}, \citenamefont {Etula}, \citenamefont {Susi},
  \citenamefont {Rinke}, \citenamefont {Laurila},\ and\ \citenamefont
  {Caro}}]{golze2022accurate}%
  \BibitemOpen
  \bibfield  {author} {\bibinfo {author} {\bibfnamefont {Dorothea}\
  \bibnamefont {Golze}}, \bibinfo {author} {\bibfnamefont {Markus}\
  \bibnamefont {Hirvensalo}}, \bibinfo {author} {\bibfnamefont {Patricia}\
  \bibnamefont {Hern{\'a}ndez-Le{\'o}n}}, \bibinfo {author} {\bibfnamefont
  {Anja}\ \bibnamefont {Aarva}}, \bibinfo {author} {\bibfnamefont {Jarkko}\
  \bibnamefont {Etula}}, \bibinfo {author} {\bibfnamefont {Toma}\ \bibnamefont
  {Susi}}, \bibinfo {author} {\bibfnamefont {Patrick}\ \bibnamefont {Rinke}},
  \bibinfo {author} {\bibfnamefont {Tomi}\ \bibnamefont {Laurila}}, \ and\
  \bibinfo {author} {\bibfnamefont {Miguel~A}\ \bibnamefont {Caro}},\
  }\bibfield  {title} {\enquote {\bibinfo {title} {Accurate computational
  prediction of core-electron binding energies in carbon-based materials: A
  machine-learning model combining density-functional theory and gw},}\
  }\href@noop {} {\bibfield  {journal} {\bibinfo  {journal} {Chemistry of
  Materials}\ }\textbf {\bibinfo {volume} {34}},\ \bibinfo {pages} {6240--6254}
  (\bibinfo {year} {2022})}\BibitemShut {NoStop}%
\bibitem [{\citenamefont {Bagus}\ \emph {et~al.}(1999)\citenamefont {Bagus},
  \citenamefont {Illas}, \citenamefont {Pacchioni},\ and\ \citenamefont
  {Parmigiani}}]{bagus1999mechanisms}%
  \BibitemOpen
  \bibfield  {author} {\bibinfo {author} {\bibfnamefont {Paul~S}\ \bibnamefont
  {Bagus}}, \bibinfo {author} {\bibfnamefont {Francesc}\ \bibnamefont {Illas}},
  \bibinfo {author} {\bibfnamefont {Gianfranco}\ \bibnamefont {Pacchioni}}, \
  and\ \bibinfo {author} {\bibfnamefont {Fulvio}\ \bibnamefont {Parmigiani}},\
  }\bibfield  {title} {\enquote {\bibinfo {title} {Mechanisms responsible for
  chemical shifts of core-level binding energies and their relationship to
  chemical bonding},}\ }\href@noop {} {\bibfield  {journal} {\bibinfo
  {journal} {Journal of electron spectroscopy and related phenomena}\ }\textbf
  {\bibinfo {volume} {100}},\ \bibinfo {pages} {215--236} (\bibinfo {year}
  {1999})}\BibitemShut {NoStop}%
\end{thebibliography}%
%\bibliography{../RHMiwa,references}

\end{document}